\documentclass{article}

\usepackage{arxiv}

\usepackage[utf8]{inputenc} 
\usepackage[T1]{fontenc}    
\usepackage{hyperref}       
\usepackage{url}            
\usepackage{booktabs}       
\usepackage{amsfonts}       
\usepackage{nicefrac}       
\usepackage{microtype}      
\usepackage{lipsum}		
\usepackage{graphicx}
\usepackage{natbib}
\usepackage{doi}

\usepackage{amssymb}
\usepackage{amsmath}

\usepackage{hyperref}
\usepackage{bm}
\usepackage{graphicx}

\title{On the structure of Vorticity and Turbulence Fields in a separated flow around a finite wing; analysis using Direct Numerical Simulation}

\author{ {Juan Carlos Bilbao-Ludena},~George Papadakis \\
	Department of Aeronautics, Imperial College London,\\
Exhibition Road,
London SW7 2AZ, UK.\\
}

\date{}

\renewcommand{\headeright}{}
\renewcommand{\undertitle}{}
\renewcommand{\shorttitle}{A preprint submitted to \textit{arXiv}}
\hypersetup{
pdftitle={A template for the arxiv style},
pdfsubject={q-bio.NC, q-bio.QM},
pdfauthor={David S.~Hippocampus, Elias D.~Striatum},
pdfkeywords={First keyword, Second keyword, More},
}

\begin{document}
\maketitle

\begin{abstract}
We investigate the spatial distributions and production mechanisms of vorticity and turbulent kinetic energy around a finite NACA 0018 wing with square wingtip profile at $Re_c=10^4$ and $10^{\circ}$ angle of attack with the aid of Direct Numerical Simulation (DNS). The analysis focuses on the highly inhomogeneous region around the tip and the near wake; this region is highly convoluted, strongly three-dimensional, and far from being self-similar. The flow separates close to the leading edge creating a large, open recirculation zone around the central part of the wing. In the proximity of the tip, the flow remains attached but another smaller recirculation zone forms closer to the trailing edge; this zone strongly affects the development of main wing tip vortex. The early formation mechanisms of three vortices close to the leading edge are elucidated and discussed. More specifically, we analyse the role of vortex stretching/compression and tilting, and how it affects the strength of each vortex as it approaches the trailing edge. We find that the three-dimensional flow separation at the sharp tip close to the leading edge plays an important role on the subsequent vortical flow  development on the suction side. The production of turbulent kinetic energy and Reynolds stresses is also investigated and discussed in conjunction with the identified vortex patterns. The detailed analysis of the mechanisms that sustain vorticity and turbulent kinetic energy improves our understanding of these highly three dimensional, non-equilibrium flows and can lead to better actuation methods to manipulate these flows.
\end{abstract}


\section{Introduction}
Lifting surfaces with finite aspect ratio produce persistent, stable and coherent tip vortices (\cite{spalart1998airplane}). These vortices are ubiquitous in transport and machinery applications, for example aircraft wings, maritime propellers, helicopter rotors, wind turbines, compressor and turbine blades etc. It is well known that wingtip vortices reduce the lift capabilities and increase drag (\cite{milne1973theoretical}). Furthermore, they can persist downstream of an aircraft for many chord lengths before they are dissipated, affecting traffic control operations and preventing the application of optimal scheduling procedures (\cite{spalart1998airplane, rossow1999lift}). 

The flow physics of the near wingtip field is very complex due to the highly turbulent and three-dimensional nature comprising multiple cross-flows (\cite{craft2006computational}). More specifically, wingtip flows involve roll-up of vortex sheets, shear layer instabilities and boundary layer separation (\cite{green2012fluid}). Traditional turbulence models (that employ the eddy-viscosity concept) cannot capture the effect of rotation that characterizes strongly vortical flows, and this has led to over-prediction of turbulence production in wings with rounded tip profiles (\cite{churchfield2009numerical}). On the other hand, experimental works on wingtip flows encounter the problem of vortex meandering, which can distort measurements if not detected and filtered out (\cite{devenport1996structure}). Nevertheless, the importance of this flow has led to many insightful and detailed experimental and numerical studies that characterise the mean velocity and vorticity dynamics in a wide range of airfoil geometries and angles of attack (\cite{giuni2013formation, birch2003rollup,bailey2006effects,katz1989effect}). \cite{chow1997mean} performed a very detailed experimental study of time-average quantities, that include turbulence measurements in the near-field wingtip vortex formation region; this work has been used as a benchmark for a large number of turbulence modeling studies. Related flow settings, such as the impingement of vortices on surfaces, the study of unsteady pitching behavior and vortex breakdown have also been analysed (\cite{wood2021large,novak2000turbulent, visbal2017unsteady,garmann2015interactions}). 

Several works have studied wingtip vortices in the far-field with a special focus on their persistence (\cite{craft2006computational, jacquin2002persistence}).\cite{zeman1995persistence} linked the weak decay to the fact that in the vortex core the velocity has solid-body rotation profile, resulting in nearly zero turbulence production, effectively suppressing the Reynolds shear stress generation. Recent works have investigated the three-dimensional flow around low-aspect-ratio finite wings at low Reynolds number (laminar flow) (\cite{taira2009effect,taira2009three, zhang2020formation}), focusing mainly on post-stall conditions and the effect of different sweep angles (\cite{zhang2020laminar}).  High-fidelity simulations at higher Reynolds numbers in the presence of turbulence have also been made (\cite{uzun2010simulations,lombard2016implicit, visbal2017unsteady}). In these simulations, shear layers and separated boundary layers have been reported as the leading sources of turbulence in the tip-flow (\cite{jiang2008large}). The study of \cite{garmann2017investigation} reports both instantaneous and time-average flows, and shows that the formation, separation, and entrainment of shear-layer substructures within the tip vortex feeding sheet are stationary whereas interactions with the wake are unsteady. \cite{smith2021wing} conducted a DNS study, and presented details related to the roll-up process and the behavior of mean axial velocity and mean pressure profiles at relatively low angles of attack. 

Despite the recent progress in the fundamental understanding of the general flow features, there is relatively little discussion in the literature on the mechanisms that result in the inception, growth and intensification/suppression of the different types of vortical structures, and how these mechanisms evolve in space. Likewise, there is a need to assess the interaction of vortical structures with the developed turbulence field.  The main goal of this work is therefore to conduct a thorough analysis of the main vorticity and turbulence production mechanisms in the near-field of a squared-tip finite wing, an area where the flow is far from being self-similar.  More specifically, we consider a moderate chord-based Reynolds number of $Re_c=10^4$ and we fully resolve all scales of the flow using DNS. The flow has  multi-scale character and we investigate in detail the multiple developing vortical patterns. 

The paper is organized as follows. Section \ref{sec:Numerical_Setup} presents the numerical setup and code validation for an infinite wing (at the same Reynolds number and angle of attack). This is followed by the description and computational details of the finite wing case (section \ref{sec:comp_details_wing}). Results sections \ref{section:Main_Features}, \ref{sec:origin_evolution_of_streamwise_vortices} and \ref{sec:turbulence_characteristics} present and analyse the main features of the three-dimensional velocity field, the generation and evolution mechanisms of the streamwise vortices, and the spatially developing turbulence field, respectively. We conclude in section \ref{sec:conclusions}.

\section{Numerical Setup and validation for an infinite wing}\label{sec:Numerical_Setup}

We consider the flow around a NACA 0018 wing with a square wingtip at a chord-based $Re_c=10^{4}$ and $10^{\circ}$ angle of attack (AoA). This wing has been employed in the study of the aerodynamics of small and medium scale unmanned aircrafts and its thickness allows to fit equipment within the compact size of the vehicle (\cite{crivellini2014spalart}). This airfoil has also been employed in wind turbines (\cite{mohamed2020numerical}) where it has been reported to improve blade stiffness and loading (\cite{timmer2008two}). The selected Re number and AoA  correspond to parts of the operational envelope of UAVs (\cite{taira2009three}). 

For the numerical simulations, we use the in-house code PANTARHEI that solves the incompressible Navier-Stokes equations using the fractional step method. The equations are discretised using the Finite Volume approach, with a second order central discretization scheme in space (for both convection and viscous terms) and a 3rd order backward scheme in time. The orthogonal viscous terms are treated implicitly, while the non-orthogonal viscous and the  convective terms are treated explicitly and marched in time with a 3rd order extrapolation. The PETSc library is employed for parallelization (\cite{petsc-efficient}). The algebraic multigrid preconditioner BoomerAMG, implemented in the Hypre library (\cite{falgout2002hypre}), is used to accelerate the solution of the linear systems resulting from the discretization process. The code has been used extensively to simulate transitional and turbulent flows in several geometries   (\cite{xiao2019nonlinear,yao2022analysis,mikhaylov2021reconstruction, schlander2022analysis}), including flows around airfoils (\cite{thomareis2017effect,thomareis2018resolvent}).

For validation purposes, we compare our results against those of \cite{zhang2015direct} for an infinite wing. The wing is placed within a uniform flow with velocity $U_\infty$, at an angle of attack of $\alpha=10^{\circ}$ and the simulation is performed at the chord-based $Re_c=U_\infty C/\nu=10^{4}$ (same conditions as the 3D finite wing). The mesh consists of approximately 10 million cells distributed across 32 planes in the spanwise direction with domain length and boundary conditions set according to \cite{zhang2015direct}. The simulation was initially run for $35C/U_\infty$, a period sufficient for the flow to reach statistically steady state. It was then restarted and time-average statistics were collected over a period of $18C/U_\infty$.

\begin{figure*}[h!]
\centering
	\includegraphics[width=0.7\linewidth]{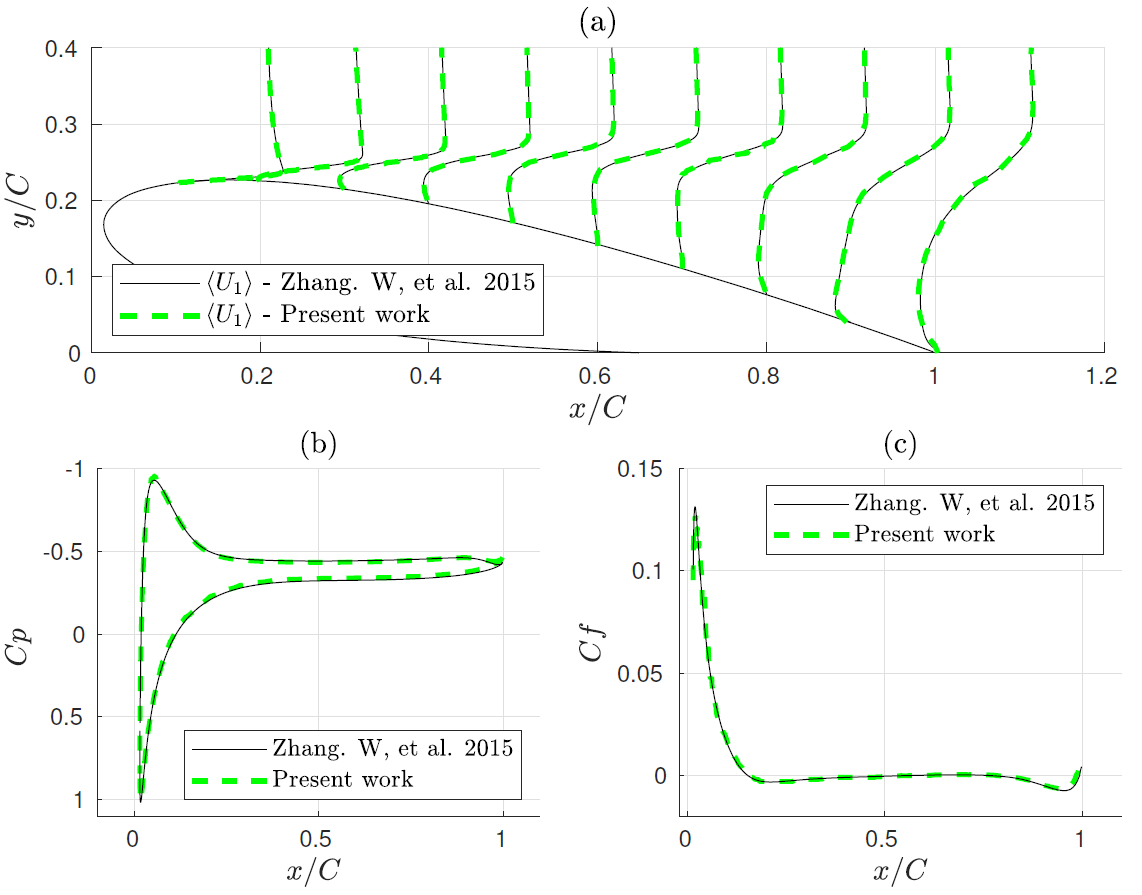}
\caption{\label{fig:mean_velocity_profiles} Comparison of time- and spanwise-averaged  (a) streamwise velocity profiles at equally spaced intervals $x/C = 0.1, 0.2, \dots, 1.0$. $(b)$  $C_p$ distribution and $(c)$ $C_f$ distribution. Present results are compared with those of \cite{zhang2015direct}. }
\end{figure*}

Fig.~\ref{fig:mean_velocity_profiles}(a) presents comparison of time- and spanwise averaged profiles of streamwise velocity at different $x/C$ locations. There is very good agreement with the results from the literature. Panel (b) shows the $C_p$ distribution, also time and spanwise-averaged. Separation occurs at $x/C \approx 0.2$ and the recirculation zone on the suction side creates a  $C_p$ plateau that extends almost until the trailing edge (note a slight increase close to $x/C=1$). The distribution of the skin friction coefficient $C_f$ on the suction side is shown in panel (c). It reaches a peak due to flow acceleration around the leading edge and quickly decays due to the adverse pressure gradient, finally reaching the value of zero, which marks the beginning of the flow separation region. Within the recirculation region, $C_f$ has very low values, as can be also detected from the shape of the velocity profiles of panel (a). The good matching with the literature data allows us to proceed with the simulation of the finite wing.

\section{Computational details for the simulation of the finite wing}\label{sec:comp_details_wing}

\begin{figure*}[h!]
\centering
\includegraphics[width=0.85\linewidth]{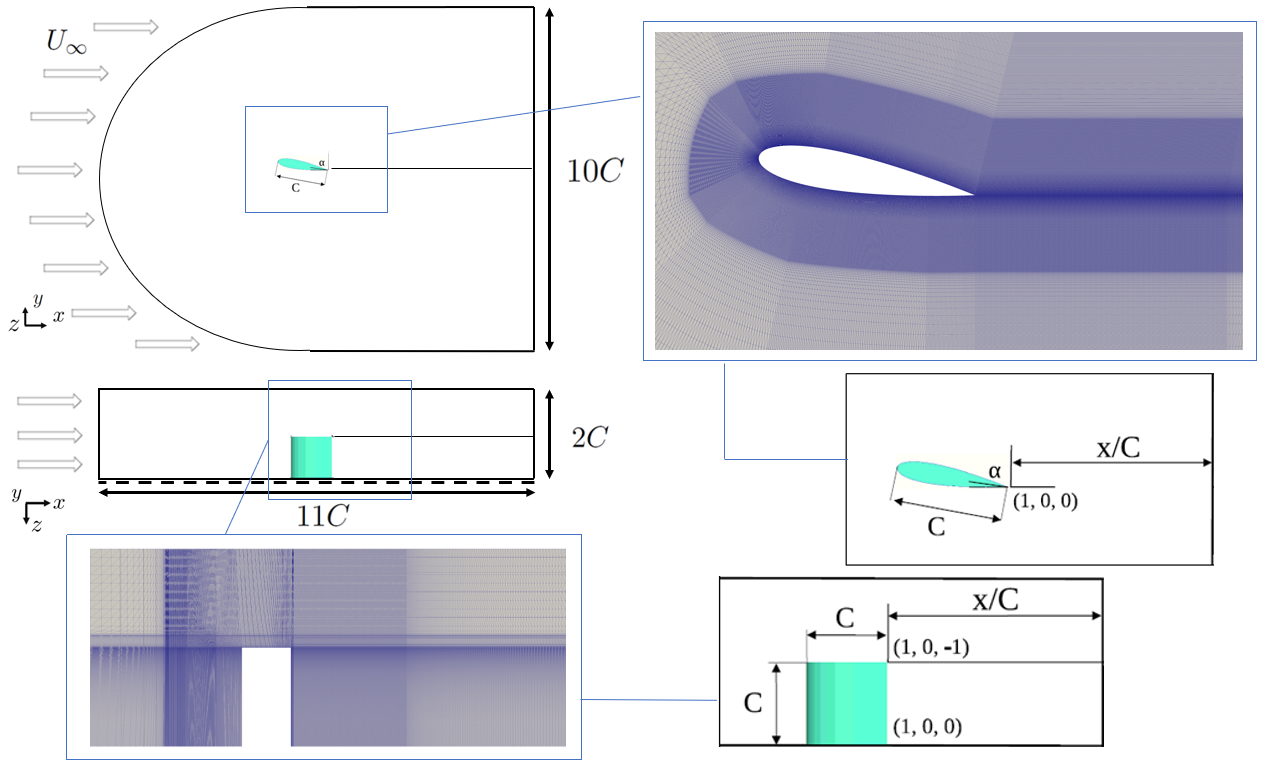}
	\caption{\label{fig:mesh_and_computational_domain_iteration_65M_3} Computational domain (front and top views) and mesh layout for the finite wing simulation. The trailing edge is located at $x/C=1$, $y/C=0$ and extends in the spanwise direction $z/C=-1 \dots 0$. Only half of the wing is simulated (and shown).}
\end{figure*}

The dimensions of the computational domain used for the finite aspect ratio wing with square tip is shown in Fig.~\ref{fig:mesh_and_computational_domain_iteration_65M_3}. A C-type body-fitted grid is wrapped around the airfoil and enables very good resolution of the near wall flow features. The mesh comprises around 63 million cells, which are clustered in the vicinity of the wing, the tip and the near wake. To reduce the computational cost, only half of the wing is simulated. The aspect ratio of the whole wing is equal to $2$ and the domain extends one chord length beyond the wingtip in the spanwise direction. Post processing of the results has shown that at a spanwise distance $0.8C$ from the wingtip, $C_p$ converges to the free-stream value.  The $(x,y,z)$  coordinates, used interchangeably with $(x_1,x_2,x_3$), are along with the streamwise, cross-stream, and spanwise directions respectively. The trailing edge is located at $x/C=1$,  $y/C=0$ and the wingtip at $z/C=-1$, refer to the bottom  right panels of Fig.~\ref{fig:mesh_and_computational_domain_iteration_65M_3}. A no-slip wall boundary condition is imposed at the wing surface, whereas the top and bottom planes (of the front view) are defined as symmetry planes. At the spanwise planes $z/C=-2$ and $0$ symmetry conditions are also imposed. The flow velocity is prescribed at the inlet,  while a convective boundary condition is employed at the outlet. 

The flow was simulated using 787 cores. After reaching steady state, the simulation continued for another $120C/U_\infty$  to collect statistics. The thickness of the cell closest to the wall has $\Delta y^{+} < 0.5$ and the grid spacings in the streamwise and spanwise directions have $\Delta x^{+}  < 3$ and $\Delta z^{+}  < 1$ respectively (the plus superscript, $^+$, indicates wall units, as usual).  This is very fine resolution, that exceeds the standard DNS requirements for wall-bounded flows $10 < \Delta x^{+}  < 20$, $\Delta y^{+}  < 1$, $5 < \Delta z^{+}  < 10$ (\cite{piomelli2002wall}). The time step was set to $0.0004$ and the maximum CFL was about 0.5.

\begin{figure*}[h!]
\centering
\includegraphics[width=0.85\linewidth]{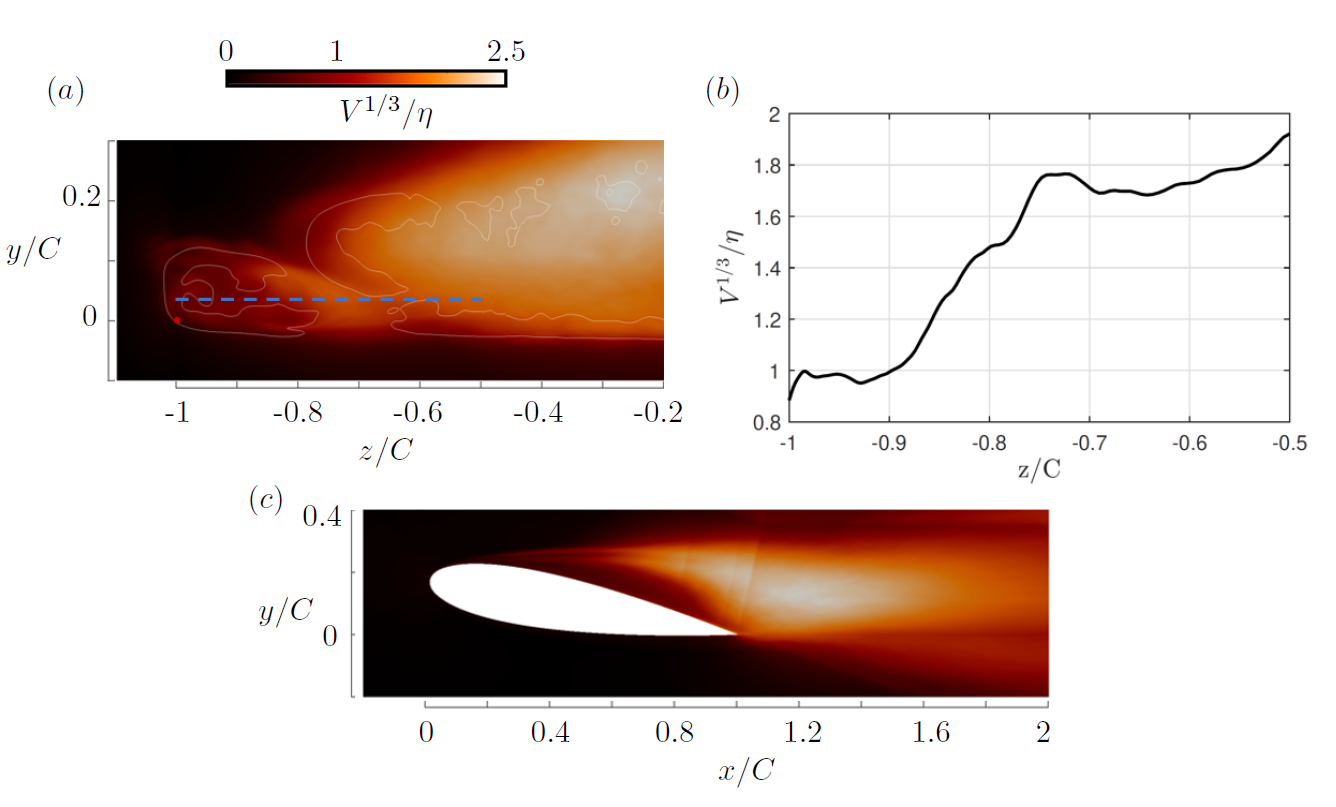}
\caption{Plots of the grid size to Kolmogorov length scale ratio, $V^{1/3}/ \eta$. (a) Contours  in the $y-z$ plane at $x/C=1.1$. The white curve represents the  $Q=5$ isoline and the red dot marks the $(z,y)$ location of the wingtip corner. (b) Plot along the dashed blue line shown in (a) at $y/C=0.04$. (c) Contour at $x-y$ plane at $z/C=-0.5$ (quarter-span).
}	\label{fig:kolmo_z_y_2}
\end{figure*}

The quality of the mesh was further checked by computing the ratio of the grid size (equal to the cubic root of the cell volume, $V^{ 1/3}$) to the Kolmogorov length scale, $\eta$. The latter was calculated from $\eta=({\nu}^{3} /\epsilon)^{1/4}$, where $\nu$ is the dynamic viscosity, $\epsilon = 2\nu \langle s_{ij} s_{ij}\rangle $ is the turbulent kinetic energy dissipation rate, and  $s_{ij}$ are the components of the fluctuating strain-rate tensor. Fig.~\ref{fig:kolmo_z_y_2}(a) shows a contour plot of this ratio in the $z-y$ plane downstream of the trailing edge; the maximum value is below 2.5. In panel (b), the values along the blue dashed line shown in panel (a) are found to be below 2. Fig.~\ref{fig:kolmo_z_y_2}(c) shows contours in the $x-y$ plane located at quarter span; again the ratio  does not exceed 2.5. The values are sufficiently small and in accordance with the recommendations of \cite{yeung1989lagrangian}. We now proceed with the presentation and analysis of the results.

\section{Main Features of the Near Wake velocity field}\label{section:Main_Features}

\begin{figure*}[h!]
\centering
\includegraphics[width=0.85\linewidth]{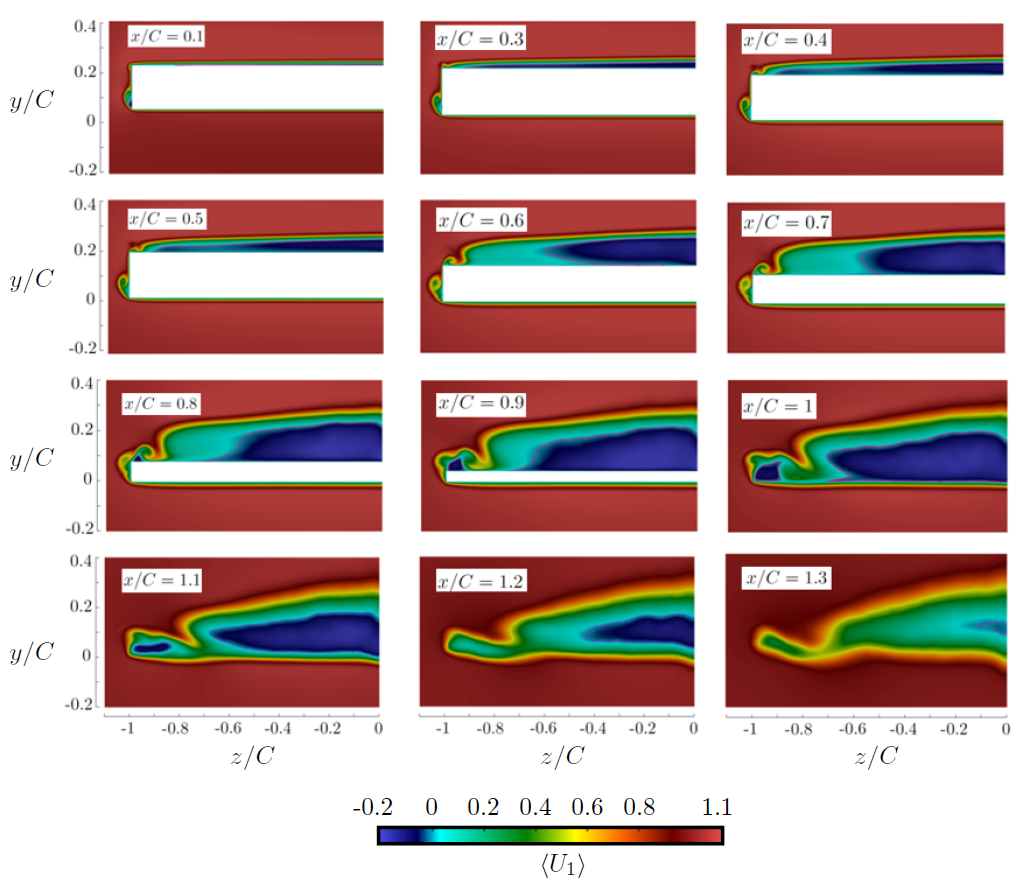}
\caption{Contours of the mean axial velocity $\langle U_1\rangle$ at equally spaced planes $x/C =0.1, \dots  1.3$. The purple-coloured contour line represents $\langle U_1\rangle=0$.
}	\label{fig:time-average-streamwise-contours-1}
\end{figure*}

Contours of the time-averaged streamwise velocity at different streamwise planes are shown in Fig.~\ref {fig:time-average-streamwise-contours-1}. The flow separates at $x/C \approx 0.2$ and does not reattach on the suction side of the wing. Close to the leading edge, the separating region occupies most of the span, but once the secondary flow in the wingtip starts to develop, the left boundary is displaced inboard. This creates a recirculating bubble with the outboard end forming an angle with the streamwise direction; this will be seen more clearly in subsequent figures. A 3D separated region with a similar trapezoidal shape has been also observed in \cite{garmann2017investigation} for a wing with a rounded tip profile. It is very interesting to see that at $x/C \approx 0.8$ another small recirculating bubble (indicated by a deep blue color) appears close to the wingtip; this bubble closes in the near wake at $x/C \approx 1.2$. Between the two recirculation regions, a jet with positive velocity appears.  As will be seen later, the shear layers in the wall-normal and spanwise directions play a very important role in the turbulence production and distribution of Reynolds stresses across the span. The flow has a wake-like behaviour, i.e\ $\langle U_1\rangle<1$ at all planes (the angular brackets indicate time-averaging). The only region where $\langle U_1\rangle >1$ is observed slightly above the wake where the velocity magnitude reaches a value of approximately 1.1.

The $Q$-criterion was used to identify the emerging coherent structures of the finite wing flow. The identification method $\lambda_2$ (\cite{jeong1995identification}) was also implemented and resulted in similar structures. $Q$ is defined as the second invariant of velocity gradient tensor $\frac{\partial U_i }{\partial x_j}$, i.e.\
$Q= 0.5 \left( \Omega_{ij} \Omega_{ij} - S_{ij} S_{ij}  \right)$, where $\Omega_{ij}$ and $S_{ij}$ are the components of the instantaneous rotation and strain-rate tensors respectively. Values $Q>0$ denote a rotation-dominated region, which can be used for vortex identification purposes.

\begin{figure*}[h!]
\centering
\includegraphics[width=0.85\linewidth]{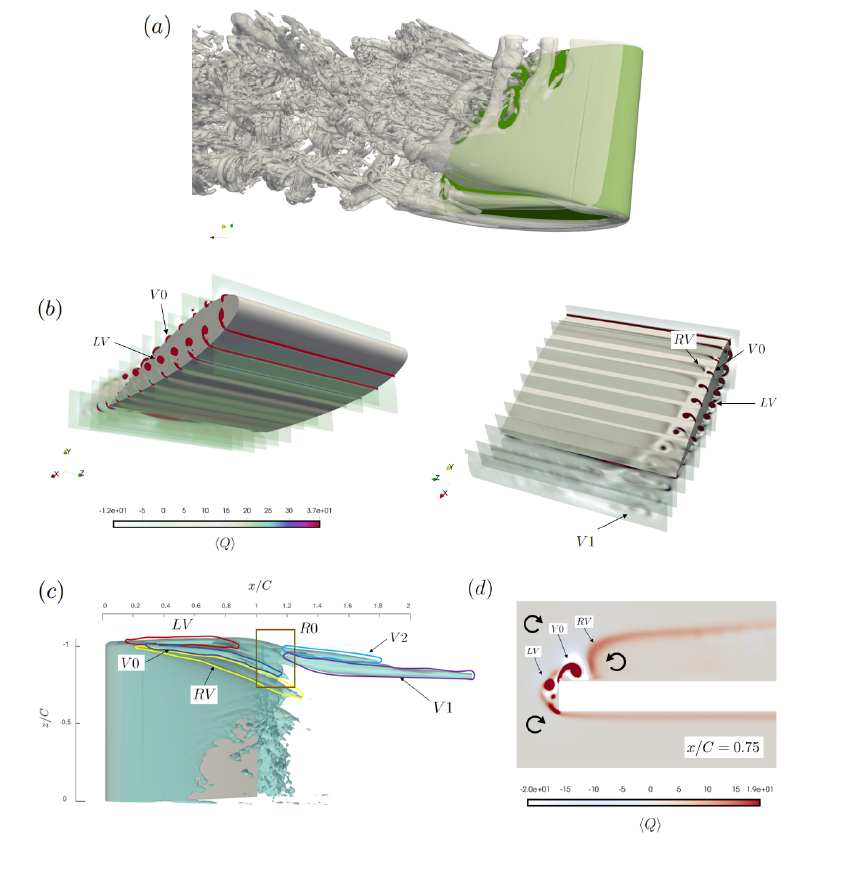}
\caption{(a) Instantaneous isosurface of $Q=5$. The wing is shown in green colour in the background. (b) Contours of the time-average $\langle Q \rangle$ at equally spaced planes $x/C = 0.1,0.2, ..., 1.0.$ with designated vortical structures. (c) Visualization and designation of vortical structures using the iso-surface $\langle Q \rangle =5$. (d) Organization of the vortex structure topology at $x/C=0.75$.}
\label{fig:Q-Criterion_all}
\end{figure*}

Fig.~\ref{fig:Q-Criterion_all}(a) shows the instantaneous isosurface $Q=5$. The laminar flow separates at the leading edge and the separated shear layer undergoes Kelvin-Helmholtz (KH) instability, leading to roll-up and the formation of KH vortices. Such vortices have also been seen in other works on infinite wings (\cite{rosti2016direct, shan2005direct,jiang2004parallel}). Further downstream the vortices break down resulting in transition to turbulence and the emergence of fine eddies in the proximity of the trailing edge. This process does not extend over the whole span though, and it is absent around the wing tip. In this area, the secondary flow from the pressure side distorts the shear layer emerging from the leading edge and prevents separation, keeping the flow laminar; there is no evidence of KH instability either. 

Visualization of isosurfaces of the time-average $Q$-criterion from different viewing angles enables us to understand better the structures that originate in the pressure area and the tip region. Fig.~\ref{fig:Q-Criterion_all}(b) shows clearly one vortex that starts at the flat tip side and grows. This vortex is designated as Left Vortex ($LV$) and arises due to the spanwise separation of the shear layer emanating from the pressure side of the wing. Near the sharp edge on the suction side, a concentrated vortex starts developing at around $x/C=0.1$, it grows and subsequently weakens as it approaches the trailing edge; we designate this vortex as $V0$. The evolution of this vortex can be visualized better in  Fig.~\ref{fig:Q-Criterion_all}(c); we see that $V0$ has decayed shortly after the trailing edge. Starting at the same initial plane as $V0$, the Right Vortex ($RV$) develops in parallel but its trace is lost at a short distance downstream of the trailing edge. 

The vortical region $R0$ is defined as the area where the separated shear layers taking part in the roll-up process amalgamate and readjust downstream of the tip. The main trailing vortex $V1$ originates at $R0$ and persists throughout the computational domain. $V2$ is another vortex that originates also from $R0$ but merges further downstream with $V1$. The organization of the vortex structure topology at $x/C=0.75$ is shown in  Fig.~\ref{fig:Q-Criterion_all}(d) and shows good agreement with visualizations reported in experimental (\cite{katz1989effect,bailey2006effects, giuni2013formation,birch2003rollup}) as well as computational studies (\cite{smith2021wing}) despite employing different $Re_c$ and geometrical setups. In the next section, we explore in more detail the origins of these vortices and their evolution. 

\section{Origins of streamwise vortical structures and their evolution due to vortex stretching/tilting mechanisms} \label{sec:origin_evolution_of_streamwise_vortices}
\subsection{Generation of Streamwise structures \label{sec:generation_of_streamwise_vortices}}

\begin{figure*}[h!]
\centering
\includegraphics[width=0.6\linewidth]{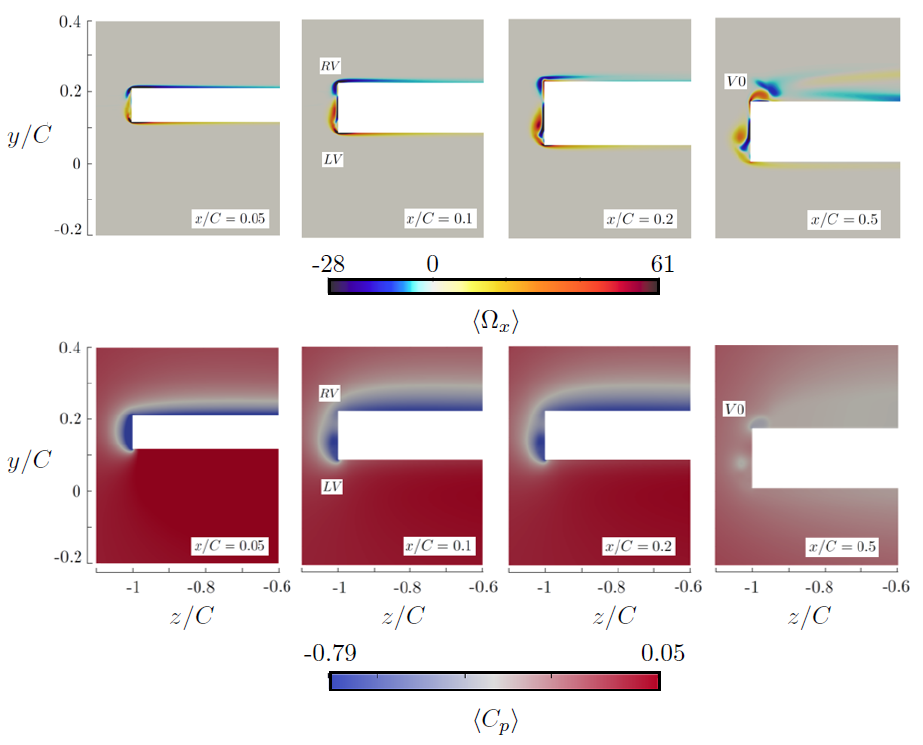}
\caption{Contours of time-averaged streamwise vorticity (top row) and pressure coefficient (bottom row) at different $x$ planes, highlighting the early formation of vortical structures $RV$, $LV$ and $V0$.
}	\label{fig:BVF_early_formation}
\end{figure*}

We start by analysing the early formation of the vortices in the wing surface. Close to the leading edge the flow is laminar, transition occurs downstream, at around $x/C \approx 0.8-0.9$, see section \ref{sec:turbulence_characteristics}. Fig.~\ref{fig:BVF_early_formation} shows contours of streamwise vorticity, $\langle \Omega_x \rangle $, close to the wing tip. Starting at $x/C=0.05$, the pressure difference between the lower and upper surface induces a relatively uniform spanwise flow leaking towards the tip. This creates the expected positive $\langle \Omega_x \rangle >0$ at the pressure side, however a rather unexpected negative $ \langle \Omega_x \rangle$ appears on the suction side (indicating a spanwise velocity directed towards the tip). This plane is located very close to the leading edge, and the secondary flow has not yet reached the suction side, as can also be seen from Fig.~\ref{fig:time-average-streamwise-contours-1}.

\begin{figure*}[h!]
\centering
\includegraphics[width=0.85\linewidth]{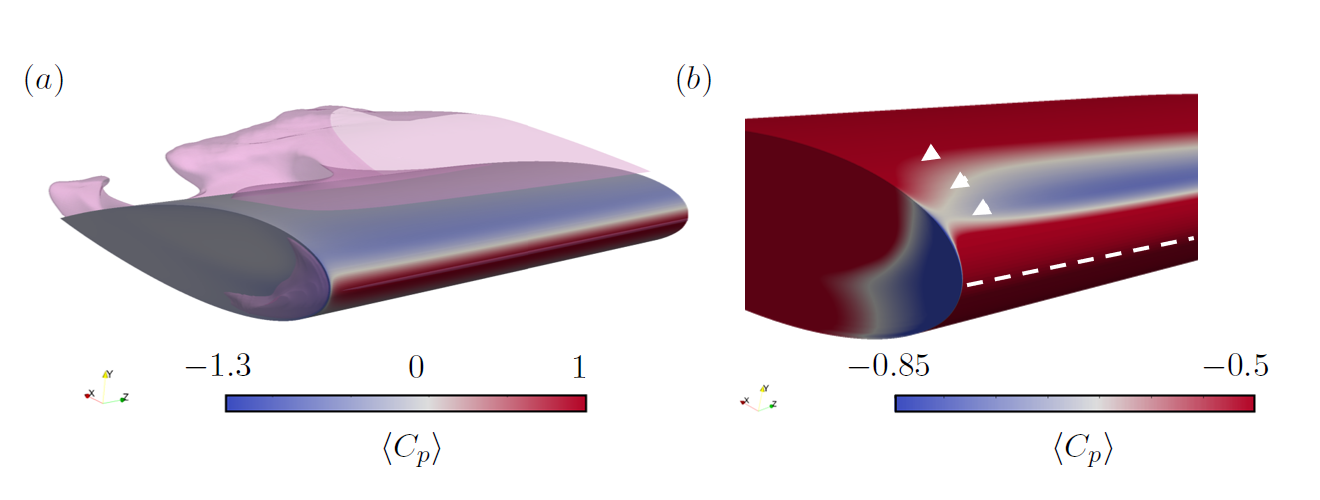}
\caption{Contours of $\langle C_p \rangle$ over the wing (a) superimposed with the $\langle U_1 \rangle=0$ isosurface, (b) zoomed-in view around the leading edge. The white dashed line marks the stagnation line across the span. The three white triangles indicate the streamwise locations $x/C=0.05$, $0.1$, and $0.2$.}	\label{fig:cp_leading_edge}
\end{figure*}

To explain this unexpected behaviour very close to the leading edge, it is instructive to consider a three-dimensional view of the distribution of the pressure coefficient  $\langle C_p \rangle=\frac{\langle P \rangle- \langle P_\infty \rangle}{0.5 \rho U_\infty^2}$ as shown in the left panel of Fig.~\ref{fig:cp_leading_edge}, where the isosurface  $\langle U_1 \rangle=0$ is also superimposed. A zoomed-in view close to the leading edge is shown in the right panel. This figure elucidates very clearly the underlying mechanism that explains the observed behaviour of streamwise vorticity. As the flow approaches and impinges on the leading edge, the pressure grows reaching $\langle C_p \rangle \approx 1$ as expected, see panel (a). Close to the tip however, the flow is diverted in the spanwise direction and separates around the sharp round edge, creating a small recirculation zone, as shown in panel (a). Note that underneath the recirculation zone a patch of very low pressure forms, see dark blue region in the flat tip in panel (b). As the flow accelerates around the suction side, the pressure is also reduced (see light blue strip in panel (b)), however it remains larger compared to the pressure on the flat tip side (this is corroborated by the pressure contours shown in the bottom row of figure 
\ref{fig:BVF_early_formation}). This pressure difference between the top surface and the flat tip results in negative spanwise velocity that is responsible for the negative $ \langle \Omega_x \rangle$. Essentially the flow is sucked towards the low pressure region forming on the tip surface. Note that the mechanism that creates this secondary motion is the pressure difference between top surface and the tip, and not between the top and bottom surfaces as in the standard lifting line theory. It is expected that the same mechanism will be also at play for rounded tips, depending on the radii of curvature in the x-z and x-y planes. If for example, the acceleration around the rounded tip in the x-z plane is faster compared to that in the x-y plane, lower pressure will appear on the surface of the tip resulting in a similar streamwise vorticity development as observed in \cite{garmann2017investigation}.

The subsequent evolution of spanwise vorticity is shown in figure  Fig.~\ref{fig:BVF_early_formation}. At $x/C=0.1$, the positive streamwise vorticity that was generated at the pressure side starts moving towards the suction side, creating the $LV$ vortex, see also panel (b) of Fig.~\ref{fig:Q-Criterion_all}. Further downstream, starting at $x/c=0.2$, the flow starts to separate also at the top tip corner, providing the earliest signs of $V0$ vortex, which is more clearly visible at $x/C=0.5$. This vortex grows underneath the shear layer with negative $\Omega_x$ that is already present in the suction side, and displaces it in the inboard direction, thus creating a bent $RV$ shear layer. The three vortices, $LV$, $V0$, $RV$ leave their footprint very clearly in the $y-z$ plane, as can be seen from panel (d) of Fig.~\ref{fig:Q-Criterion_all}. This analysis shows that the separating shear layer from the leading edge (see Fig.~\ref{fig:time-average-streamwise-contours-1}) acquired negative streamwise vorticity very early on (due to the developed pressure distribution as already mentioned), and vortex $RV$ is the result of intrusion and bending of this shear layer by $V0$. 

\begin{figure*}[h!]
\centering
\includegraphics[width=0.7\linewidth]{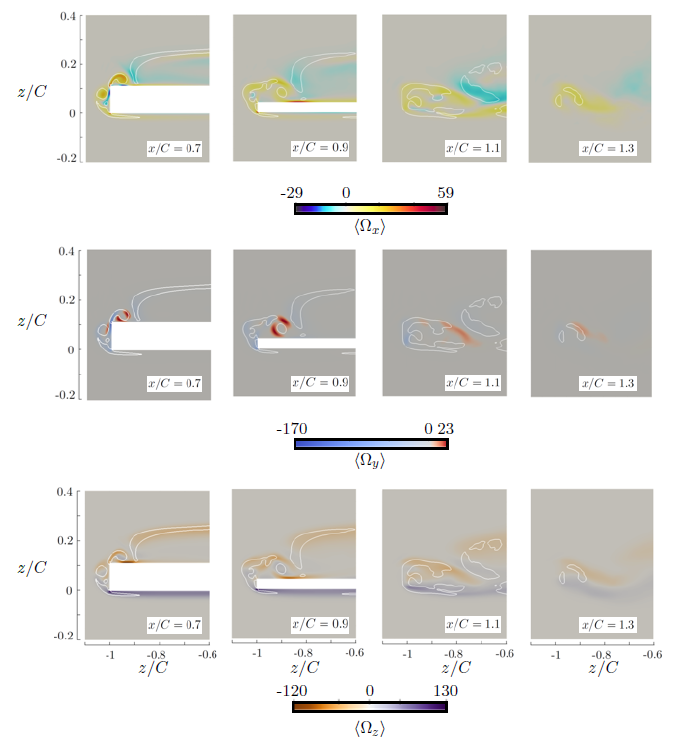}
\caption{\label{fig:Vortical_component_magnitudes} Contour plots of time-averaged vorticity components (streamwise at top row; wall-normal at middle row; spanwise at bottom row)  at different $x/C$ planes. The isoline $\langle Q\rangle =7$ is indicated in white colour. 
}
\end{figure*}

Fig.~\ref{fig:Vortical_component_magnitudes} shows contour plots of $ \langle \Omega_{x} \rangle $, $\langle \Omega_{y} \rangle $ and $ \langle \Omega_{z} \rangle $ at different streamwise planes further downstream. This figure clearly highlights the multiple and diverse vortical structures formed around the sharp corners of the wingtip. The $ \langle \Omega_{x} \rangle $ component maintains its coherence at $x/c=0.7$, and all the three aforementioned vortical structures are visible. Further downstream, at $x/C=0.9$, $V0$ and $LV$ (that have the same sense of rotation) seem to merge, leading to a large vortex structure, which retains its coherence at  $x/c=1.1$. It is interesting to observe the formation of a thin layer of strong positive $ \langle \Omega_{x} \rangle$ close to the wall, in the gap between $V0$ and $RV$ at $x/C=0.9$. It is difficult to explain the origin of this layer, but probably it is related to the spanwise pressure gradient generated by the unequal vorticities of $V0$ and $RV$.    

The wall-normal component, $ \langle \Omega_{y} \rangle =\frac{\partial \langle U_1 \rangle }{\partial x_3}-\frac{\partial \langle  U_3 \rangle  }{\partial x_1}$, which encodes the spanwise inhomogeneity of $\langle U_1 \rangle $ velocity, peaks at two regions located in the boundaries between the three vortices (see contours at $x/C=0.7$ and $0.9$). Further downstream, only one peak remains, while in the wake $ \langle \Omega_{y} \rangle$ decays. 

The stronger vorticity component as expected is $ \langle \Omega_{z} \rangle $, but its detailed structure is significantly distorted by the strongly three-dimensional nature close to the wing tip. Note that both positive and negative values appear at $x/C=0.7-1.3$ in the suction side. The positive ones are due to the recirculation zone close to the tip shown in Fig.~\ref{fig:time-average-streamwise-contours-1}. In the near wake, at $x/c=1.3$, the shape is dictated by the two wakes from the top and bottom sides, and the strong three-dimensionality has subsided. 

\subsection{The role of vortex stretching and tilting on the evolution of vortical structures - Early stage}

We now investigate how the three vortices identified earlier sustain or lose their strength. To this end, we employ the evolution equation of the time-average vorticity component $\langle \Omega_{i}\rangle $, which takes the form  (\cite{tennekes2018first}):
\begin{equation}
	\langle U_j \rangle  \frac{\partial \langle \Omega_{i}\rangle }{\partial x_j} + \langle u_j \frac{\partial \omega_{i} }{\partial x_j}\rangle =  \langle \omega_{j}s_{ij}\rangle +\langle \Omega_{j}\rangle \langle S_{ij} \rangle  + \nu  \frac{\partial^{2} \langle \Omega_i \rangle }{\partial x_{j}^2} ,
	\label{eq:vorticity_transport_equation}
\end{equation}
where lower case variables denote fluctuations, i.e.\ the Reynolds decomposition is $\Omega_{i} =\langle \Omega_{i}\rangle + \omega_i$. 
Vorticity can be amplified or suppressed due to the vortex stretching/tilting term $\langle \omega_{j}s_{ij}\rangle +\langle \Omega_{j}\rangle \langle S_{ij} \rangle$  (\cite{wu2015vortical}). In the following, we will investigate this term in more detail. In particular, we consider only the streamwise vorticity  $\langle \Omega_{x}\rangle $ and the vorticity intensification $\langle \Omega_{j}\rangle \langle S_{xj} \rangle$, which can be decomposed into vortex stretching (or axial elongation) and tilting terms,
\begin{equation}
	\langle \Omega_{j}\rangle S_{xj}=\underbrace{\langle \Omega_{x}\rangle \langle S_{xx} \rangle}_{\text{vortex stretching}}+\underbrace{\langle \Omega_{y}\rangle \langle S_{xy} \rangle+\langle \Omega_{z}\rangle \langle S_{xz} \rangle }_{\text{vortex tilting}}.
\end{equation}
Stretching and tilting represent features of vorticity kinematics which act only in 3D flows. From  conservation of angular momentum, a vortex tube undergoing mean stretching, i.e.\ $\langle S_{xx} \rangle=\frac{\partial \langle U_1 \rangle}{\partial x_1}>0$, will become more narrow and have increased streamwise vorticity. On the other hand, under mean compression i.e.\ $\langle S_{xx} \rangle <0$, the vorticity tube will grow thicker, resulting in a reduced rotational rate (\cite{wu2015vortical}). 

\subsubsection{Vortex V0 emerging from the top sharp edge of the wing tip}

\begin{figure*}[h!]
\centering
\includegraphics[width=0.9\linewidth]{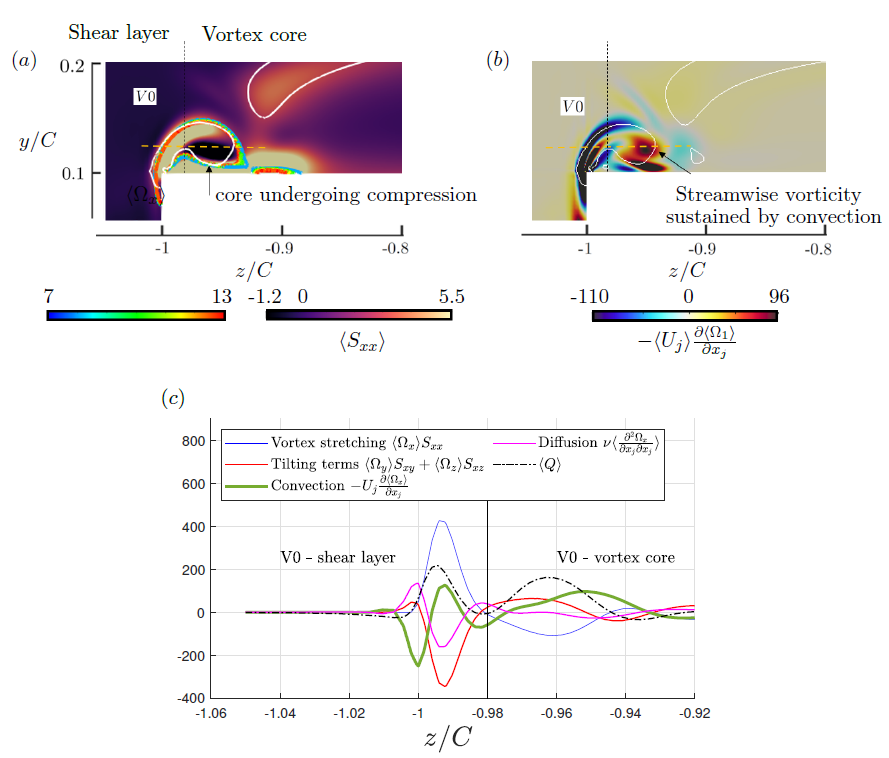}
\caption{\label{fig:V0_06vorticity_transport_} (a) Axial elongation rate (filled purple contours) superposed with streamwise vorticity (blue/red) line contours, (b) Convection term (filled contours) and (c) budgets of the vorticity transport equation along the horizontal yellow dashed line shown in panels (a) and (b). The white isoline depicts $\langle Q \rangle = 5$. Results at plane  $x/C=0.6$. 
}
\end{figure*}

We start with $V0$ and analyse the different terms of the vorticity transport equation at $x/C=0.6$, where the flow is still laminar. Fig.~\ref{fig:V0_06vorticity_transport_}(a) shows (filled) contours of axial elongation superposed with (blue/red) line contours of $\langle \Omega_x \rangle$.  We also plot in white line the contour of $\langle Q \rangle = 5$. The colored line contours of vorticity and $Q$-criterion match well and demarcate the boundaries of V0. It can be seen that this vortical structure consists of a separating shear layer emanating from the sharp top edge of the tip and a vortex core. In the shear layer the axial strain rate  $\langle S_{xx} \rangle$ is positive indicating elongation, while in the core it is negative. Axial vorticity is therefore intensified in the shear layer and subsequently feeds the core.  This mechanism is confirmed in panel (c) which shows the variation of the stretching and tilting terms along a horizontal line that crosses V0 (yellow dashed line in panel (a)).  The vertical line in the panel (c) demarcates the areas of positive and negative vortex stretching. Note that the tilting terms have opposite signs compared to the stretching term, but the latter is stronger in the shear layer region, resulting in net positive production. In the core region however, the net production is negative, leading to the eventual decay of V0. Contours of the convection term  $-\langle U_j \rangle  \frac{\partial \langle \Omega_{1}\rangle }{\partial x_j}$ are shown in panel (b). This term 
takes positive values in the middle of the shear layer and the core (see also panel c), indicating the transport of diminishing $\langle \Omega_x \rangle$ downstream.

\begin{figure*}[h!]
\centering
\includegraphics[width=0.7\linewidth]{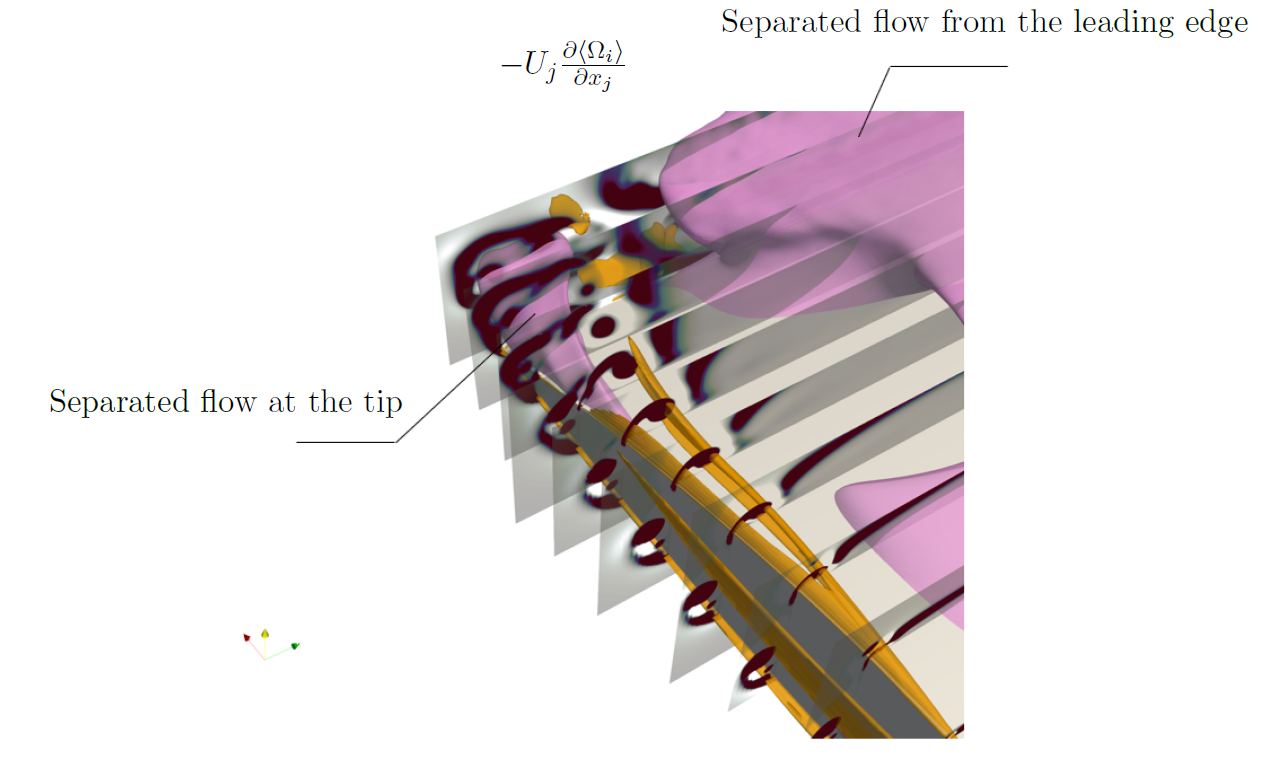}
\caption{\label{fig:V0_06vorticity_convection} Isosurface of vorticity convection $-U_j \frac{\partial \langle \Omega_{1}\rangle }{\partial x_j}=78$ (orange colour).  Vortex V0 is visualised by $\langle Q \rangle$ criterion (same values as in Fig.~\ref{fig:Q-Criterion_all}(b)) at equally spaced planes $x/C=0.4-1.2$. The pink isosurface represents $\langle {U_1} \rangle=0$ and encloses the two separated flow regions (one that starts in the leading edge, and the other in the trailing edge, around the tip). The vorticity convection isosurface value corresponds to $\approx52\%$ of the positive maximum value found within the core of V0 at the z-y plane located at $x/C=0.6$.
}
\end{figure*}

Fig.~\ref{fig:V0_06vorticity_convection} displays an isosurface of $-\langle U_j \rangle  \frac{\partial \langle \Omega_{1}\rangle }{\partial x_j}$ with positive value. The isosurface consists of two legs that correspond to the two areas mentioned above. The outboard leg terminates at the point where the separated flow at the tip (shown in pink colour) starts to form. 
The inboard leg continues slightly further downstream. Note the disintegration and amalgamation of V0 with LV close to the trailing edge. 

\begin{figure*}[h!]
\centering
\includegraphics[width=1\linewidth]{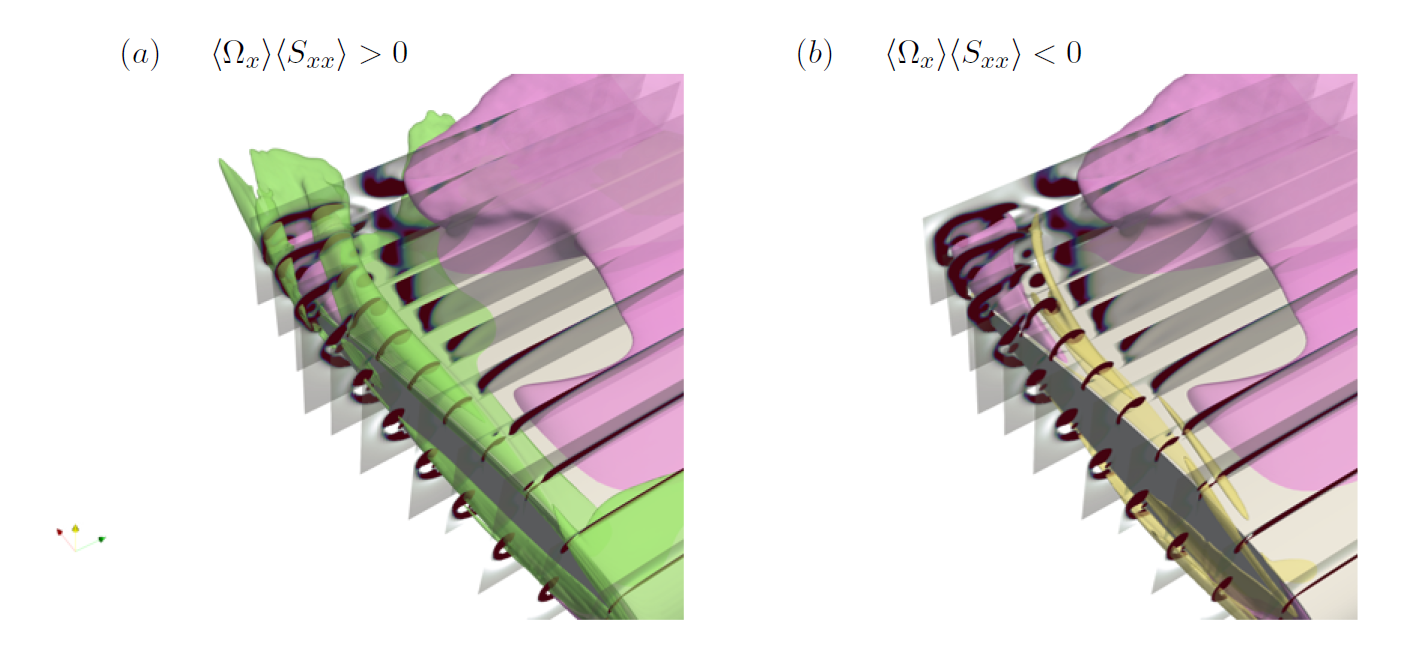}
\caption{\label{fig:V0_dynamics_pressure_gradient} Isosurfaces of (a) vorticity stretching rate  $\langle \Omega_x \rangle \langle S_{xx} \rangle=14$ (in green colour) (b) vorticity compression rate $\langle \Omega_x \rangle  \langle S_{xx} \rangle=-30$ (in yellow colour). The pink isosurface represents $\langle {U_1} \rangle=0$. Results are shown at equally spaced planes $x/C=0.2-1.2$. The isosurface values of the vortex stretching/compression rate correspond to $(a)$ $\approx32\%$  of the positive maximum value found within V1 at $x/C=1.2$ and $(b)$ $\approx44\%$ of the minimum negative value within the core of V0 at the z-y plane located at $x/C=0.85$.
}
\end{figure*}

Regions undergoing strong vorticity stretching or compression over the wing surface are shown in Fig.~\ref{fig:V0_dynamics_pressure_gradient}. 
The strong three-dimensionality of the flow is once again evident. The feeding shear layer emanating from the tip is a continuous source of positive vorticity to the core of V0 through stretching as can be seen from panel (a). This mechanism, first identified in Fig.~\ref{fig:V0_06vorticity_transport_} for a particular plane, is found to persist along the downstream direction but is distorted due to the presence of the separated region at the tip. In particular, the vortex stretching region lifts above the recirculating bubble. The region of vortex compression, shown in panel (b), is more compact and is localised around the core of vortex V0. As the latter is deflected inboard due to the recirculation zone around the tip, so does the compression region. Note that the presence of the recirculation zone severs the ties between the feeding shear and the core of V0. This has important implications for this vortex. While the compression mechanism is present from the early stages of V0 and persists throughout, once the tip recirculation bubble forms, the feeding shear can no longer provide a source of vorticity through stretching, leading to the eventual demise of V0.  

\begin{figure*}[h!]
\centering
\includegraphics[width=0.8\linewidth]{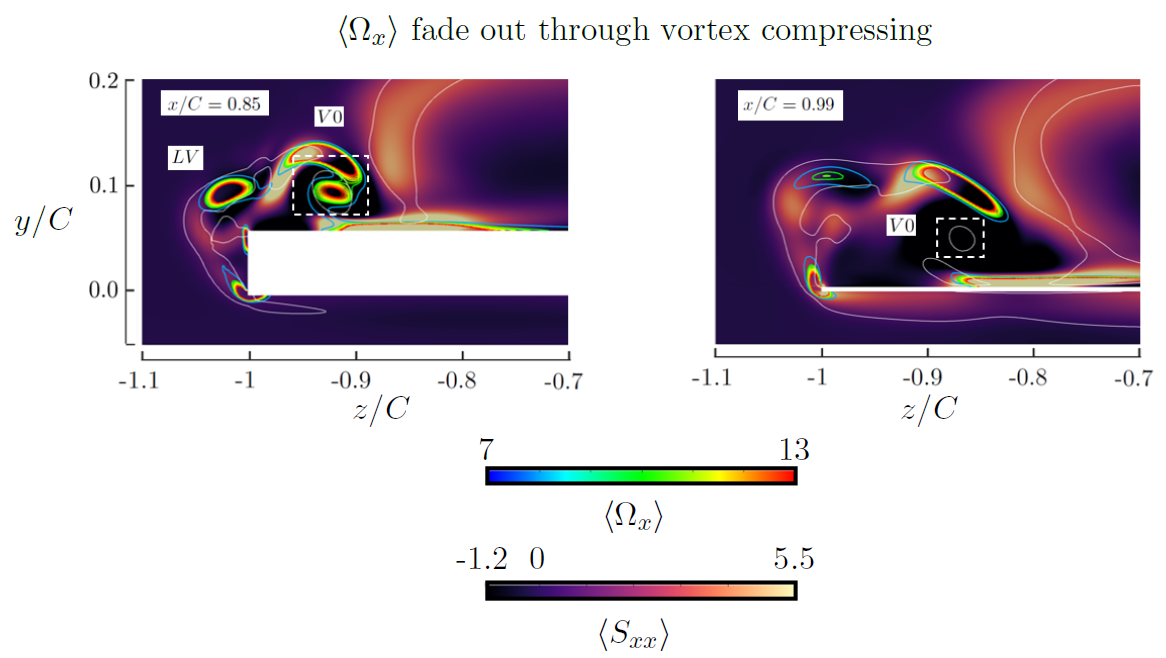}
\caption{\label{fig:V0_vortex_production_stretching} Coloured (filled) contours of elongation rate superposed with coloured (line) contours of $\langle \Omega_x \rangle$ at two planes $x/C=0.85$ (left) and $x/C=0.99$ (right). The white isoline depicts $\langle Q \rangle = 5$.   
}
\end{figure*}

This can be attested in  Fig.~\ref{fig:V0_vortex_production_stretching} that shows the vortical structures at two planes close to the trailing edge, at $x/C=0.85$ and $0.99$. The mean axial strain rate is displayed in the background with filled contours, on which coloured isolines of $\langle \Omega_x \rangle$ are superposed together with the $\langle Q \rangle = 5$ white isoline. Note how the arc-shaped shear layer originating from the pressure side is further lifted away from the surface of the airfoil as $x/C$ grows from $0.85$ to $0.99$. This is due to the presence of the recirculating bubble, as already mentioned. Vortex V0, shown here enclosed within a rectangle with dashed-lined perimeter, is disconnected from the feeding shear, thus losing its feeding source. Although it is still a coherent vortex in these two planes, shortly downstream it completely disintegrates and merges with the wake in the periphery of the trailing edge. This figure further corroborates the fact that compression effects play the dominant role in the annihilation of V0. On the other hand, the lifted shear layer carries strong streamwise vorticity (in the form of LV) which, after some readjustment, will eventually develop into the trailing vortex V1. This shear layer displaces the RV vortex further inboard. Note again the presence of a wall layer with strong positive streamwise vorticity. The layer has a finite extent in the spanwise direction and is confined between V0 and RV.

\subsubsection{Vortex LV}

\begin{figure*}[h!]
\centering
\includegraphics[width=0.7\linewidth]{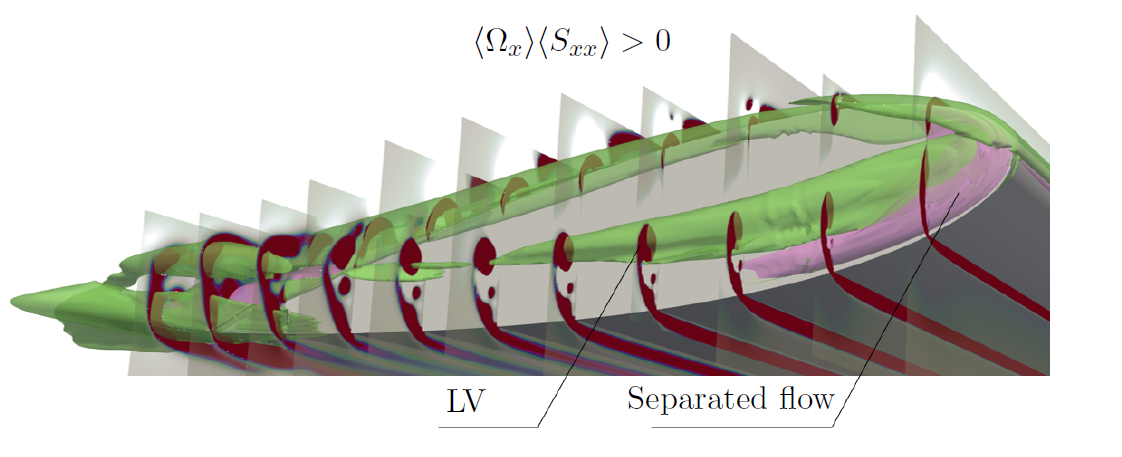}
\caption{\label{fig:LV_dynamics} Isosurface of vorticity stretching rate $\langle S_{xx}\rangle =14$ in the tip (green colour). Vortex LV is visualised by plotting the $\langle Q \rangle$ criterion (same values as in Fig.~\ref{fig:Q-Criterion_all}(b)) at equally spaced planes $x/C=0.1-1.1$. The pink isosurface represents $\langle {U_1} \rangle=0$. The isosurface $\langle S_{xx}\rangle$ value corresponds to $\approx11\%$ of the positive maximum value found within the core of LV at  the z-y plane located at $x/C=0.3$.	
}
\end{figure*}

Attention is now turned to vortex LV, which originates on the pressure side and develops along the flat tip face. Fig.~\ref{fig:LV_dynamics} shows an isosurface of the vorticity stretching rate. Note that LV starts to develop early, at around $x/C \approx 0.2$, with its core forming slightly above the aforementioned recirculation bubble that appears in the flat tip close to the leading edge. This makes physical sense; outside the recirculation zone, the flow accelerates in the streamwise direction, leading to a positive strain rate that increases the streamwise vorticity of LV. This source term however becomes weaker further downstream. As the flow develops, the core of LV is lifted up and at $x/C \approx 0.9$, it emerges on the suction side. In this region, it is amalgamated with the arch-shaped shear layer which feeds V0 with vorticity; this shear layer is found above the trailing edge recirculation pocket, as was seen more clearly in Fig.~\ref{fig:V0_dynamics_pressure_gradient}.

\begin{figure*}[h!]
\centering
\includegraphics[width=0.8\linewidth]{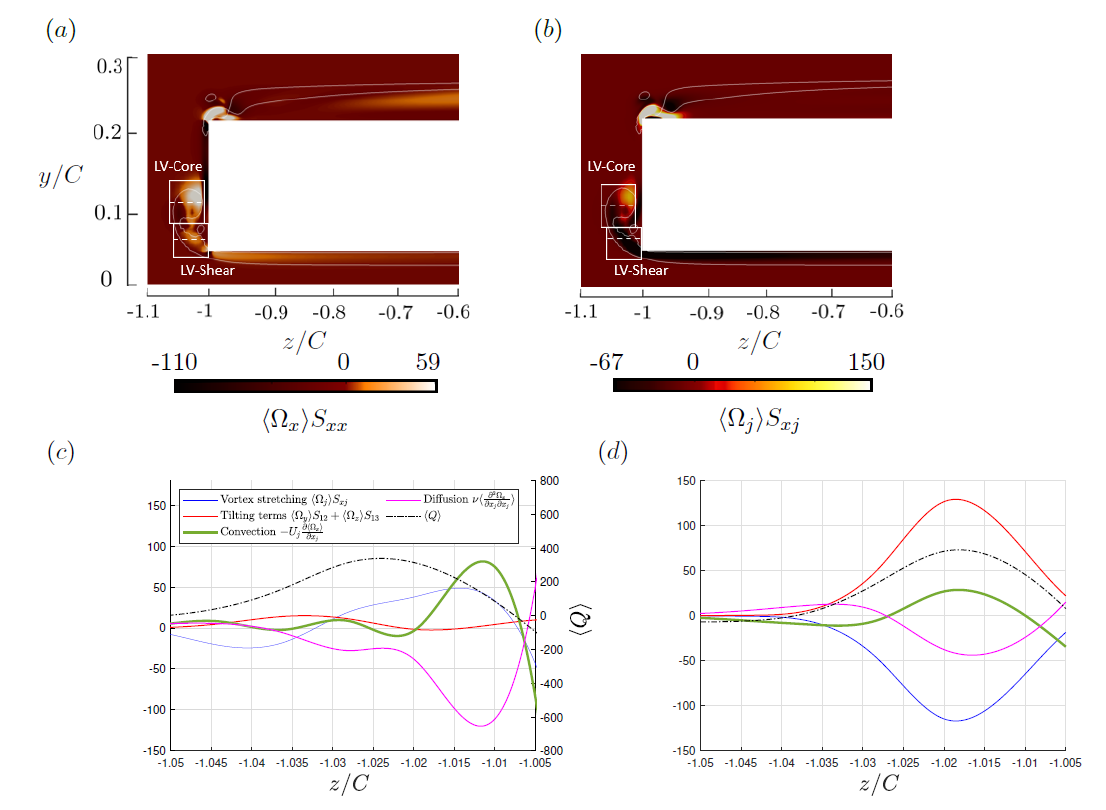}
\caption{\label{fig:LV_vortex_production_stretching} Contour plots of axial vorticity generation term (a)  stretching only and (b) stretching and tilting, for vortex LV at $x/C=0.35$. The solid white isoline depicts $\langle Q \rangle = 5$. Spanwise distribution of the budget terms of the vorticity transport equation along (c) the core and (d) the feeding shear of LV. The exact locations are marked as white dashed lines in panels (a) and (b).
}
\end{figure*}

Fig.~\ref{fig:LV_vortex_production_stretching} shows the distribution of the vorticity generation terms within the feeding shear and the core of vortex LV at $x/C=0.35$. The figure reveals the mechanism with which LV gains axial vorticity. Within the separated shear layer feeding LV, vortex tilting is the dominant positive term; this term is plotted with a solid red line in panel (d). Indeed strong spanwise vorticity is produced in the pressure side and the component $\langle \Omega_z \rangle \langle S_{13} \rangle$ of the titling term is responsible for generating $\langle \Omega_x \rangle$ in the tip. The shear layer then feeds the LV core with streamwise vorticity. In the core itself, the titling term is negligible, as shown in panel (c), and LV is sustained by a weakly positive stretching term.

\subsubsection{The counter-clockwise rotating vortex RV}

\begin{figure*}[h!]
\centering
\includegraphics[width=0.8\linewidth]{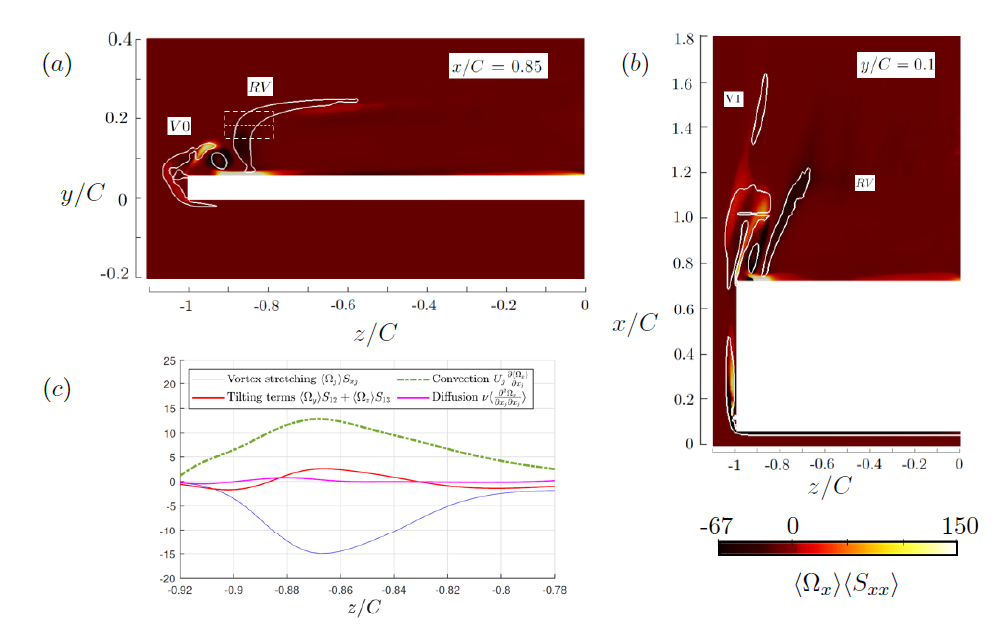}
\caption{\label{fig:RV_up_dynamics_and_vortex_production} Contours of vortex stretching rate $ \langle\Omega_x \rangle \langle S_{xx}\rangle$ at planes (a) $x/C=0.85$ and (b) $y/C=0.1$. (c) Balance of the vorticity transport equation along the dashed yellow line shown in panel (a) that crosses RV.  The solid white isolines in (a) and (b) depict $\langle Q \rangle = 5$.}
\end{figure*}

The generation mechanism of vortex RV close to the leading edge was elucidated earlier in section \ref {sec:generation_of_streamwise_vortices}. Here we investigate the mechanism that sustains RV. Fig.~\ref{fig:RV_up_dynamics_and_vortex_production} (a) shows  contours of vortex stretching rate $ \langle\Omega_x \rangle \langle S_{xx}\rangle$ at plane $x/C=0.85$. As already mentioned, RV is located to the right of V0 and is suspended at a higher wall-normal distance. The term $\langle \Omega_x \rangle \langle S_{xx} \rangle$ is negative, see colour scale at the bottom of panel (b). Since $\langle \Omega_x \rangle <0$, RV is subject to stretching and this is the mechanism that sustains it. This can also be seen from panel (c) that plots the budgets of the vorticity equation along the yellow horizontal dashed line that crosses RV, as shown in panel (a). The tilting terms play a minor role. 
Fig.~\ref{fig:RV_up_dynamics_and_vortex_production} (b) shows contours of stretching term in the $x-z$ plane with $y/C=0.1$. The plots show all vortices examined so far, and clearly display the different signs of the stretching term. Strong positive values appear for LV, and negative for V0 and RV. But V0 carries positive vorticity, RV negative, so this term has the opposite effect, it is detrimental for the former and favourable for the latter. This explains why V0 is annihilated shortly downstream of the trailing edge while RV persists, as be clearly seen in panel (b).  Eventually RV also disintegrates, but it does so further downstream, at around $x/C \approx 1.2$. 

\subsection{Emergence of the wake tip vortices V1 and V2 - Late stage}

\begin{figure*}[h!]
\centering
\includegraphics[width=0.9\linewidth]{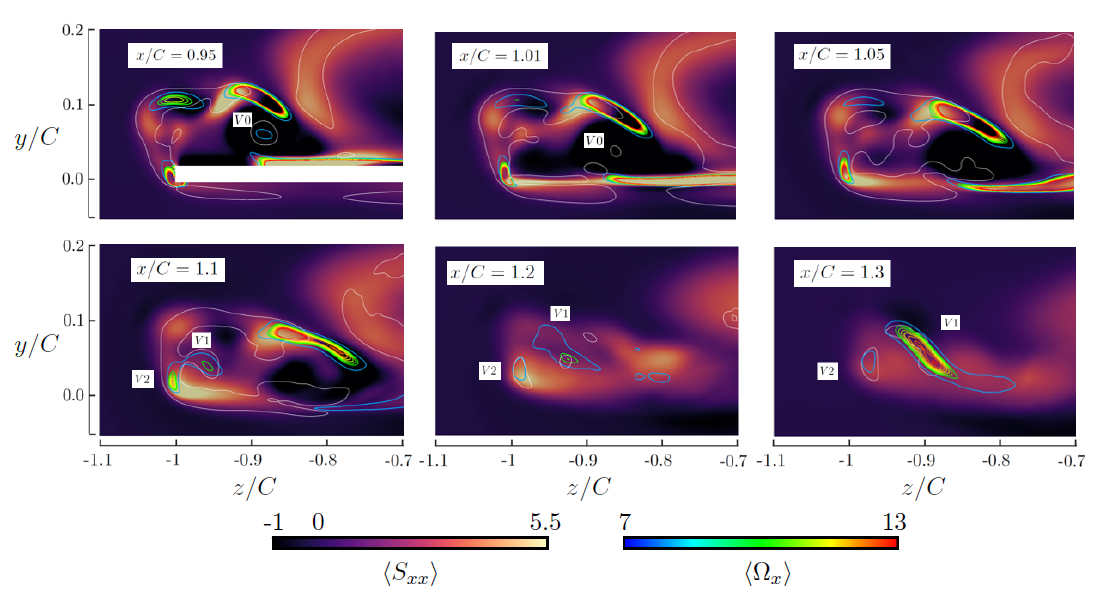}
\caption{\label{fig:Capture_origins_of_streamwise_vortices_iteration_1} Isosurfaces of vortex intensification terms,  stretching and/or tilting, superimposed on $\langle Q \rangle =5$. $(a)$ Stretching term  $\langle \Omega_x \rangle \langle S_{xx}\rangle =14$, $(b)$ Tilting terms $ \langle \Omega_y \rangle \langle S_{xy} \rangle+\langle \Omega_z \rangle \langle  S_{xz} \rangle=14$, $(c)$ Sum of all terms, $ \langle \Omega_j \rangle \langle  S_{xj} \rangle=12$. $(d)$ $\langle \Omega_{x}\rangle=8$. The isosurface values correspond to $(a)$ $\approx32\%$, $(b)$ $\approx67\%$, $(c)$ $\approx54\%$ and $(d)$ $\approx72\%$ of their corresponding positive maximum values within V1  in the z-y planes located at $x/C=1.2$ (at $x/C=1.3$ for $(b)$).}
\end{figure*}

\begin{figure*}[h!]
\centering
\includegraphics[width=0.9\linewidth]{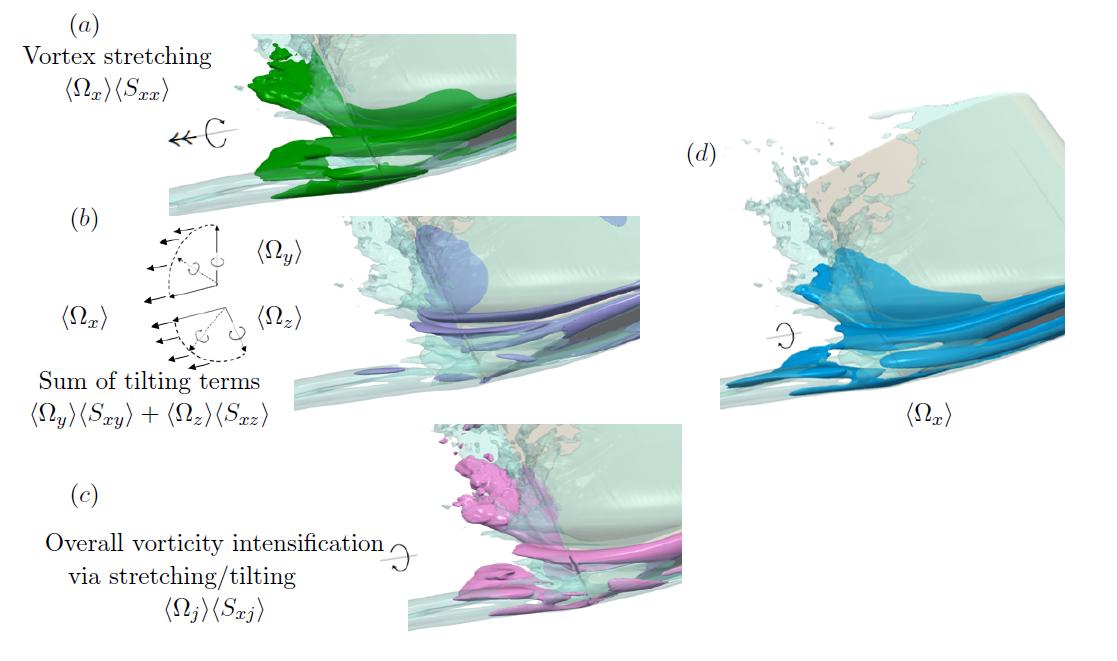}
\caption{\label{fig:Vortex_stretching_component_evolution_iteration_1} Emergence of the wake tip vortices V1 and V2. Contours (filled) of streamwise strain rate $\langle S_{xx}\rangle$ superposed to coloured isolines of positive $\langle \Omega_x \rangle$ at different $x/C$  planes. The solid white line indicates $\langle Q \rangle = 5$.}
\end{figure*}

Fig.~\ref{fig:Vortex_stretching_component_evolution_iteration_1} presents (filled) contours of streamwise strain rate $\langle S_{xx} \rangle$ superposed with isolines of positive vorticity $\langle \Omega_x \rangle$. The arch-like structure that was formed from the secondary flow around the recirculation zone at the tip persists up to about  $x/C=1.1$. This structure carries two regions of concentrated vorticity, located at approximately the same wall normal distance, $y/C=0.1$, and at spanwise locations $z/C=-1$ and $-0.9$. The former patch of vorticity quickly disappears, because it is located within a region of compression (dark contours), but the latter patch grows due to stretching (note the yellow-coloured spot of high acceleration at $z/C=0.9$ in planes $x/C=0.95-1.1$). Eventually this patch will evolve to form the tip vortex V1 that will persist in the wake, see plane  $x/C=1.3$. The core of this vortex is located slightly inboard with respect to the tip, at $z/C\approx -0.9$. This agrees with the theoretical analysis of Milne-Thomson (\cite{milne1973theoretical}). We conjecture that the high strain rate that leads to the intensification of V1 is due to the recovery of the streamwise velocity downstream of the  recirculation bubble that forms around the tip of the trailing edge.  
It is also interesting to observe that the near wall vorticity patch (that forms between V0 and RV) is immersed in a region of compression shortly downstream of the trailing edge and quickly disappears (it is detected until $x/C \approx 1.1$).  Vortex V2 starts to be visible in the lower tip corner in planes $x/C=0.95-1.05$ and undergoes stretching in all planes. It eventually merges with V1, as shown in figure \ref{fig:Q-Criterion_all}(c).

Isosurfaces of the vortex intensification terms,  stretching and/or tilting, are shown in Fig.~\ref{fig:Capture_origins_of_streamwise_vortices_iteration_1}. Panel (a) depicts the isosurface of the stretching term which occupies a large part of the wing tip region but has a fine extent downstream. Comparison with the $\langle \Omega_{x}\rangle$ isosurface, shown in panel (d), indicates a high degree of correlation. The tilting terms, panel (b), are more sporadic (we have used the same isosurface value for the stretching and tilting terms in order to make the comparison meaningful).  Finally the total intensification term shown in panel (c), indicates a high degree of resemblance to the stretching term. This  indicates that the titling term is very important in the shear layer feeding LV with vorticity, see figure \ref{fig:LV_vortex_production_stretching}), but once this function is completed, the stretching term takes over and is the dominant factor in enhancing or suppressing the local vorticity. In this sense, the mechanism is consistent with the standard lifting line theory (\cite{milne1973theoretical}).

\begin{figure*}[h!]
\centering
\includegraphics[width=0.85\linewidth]{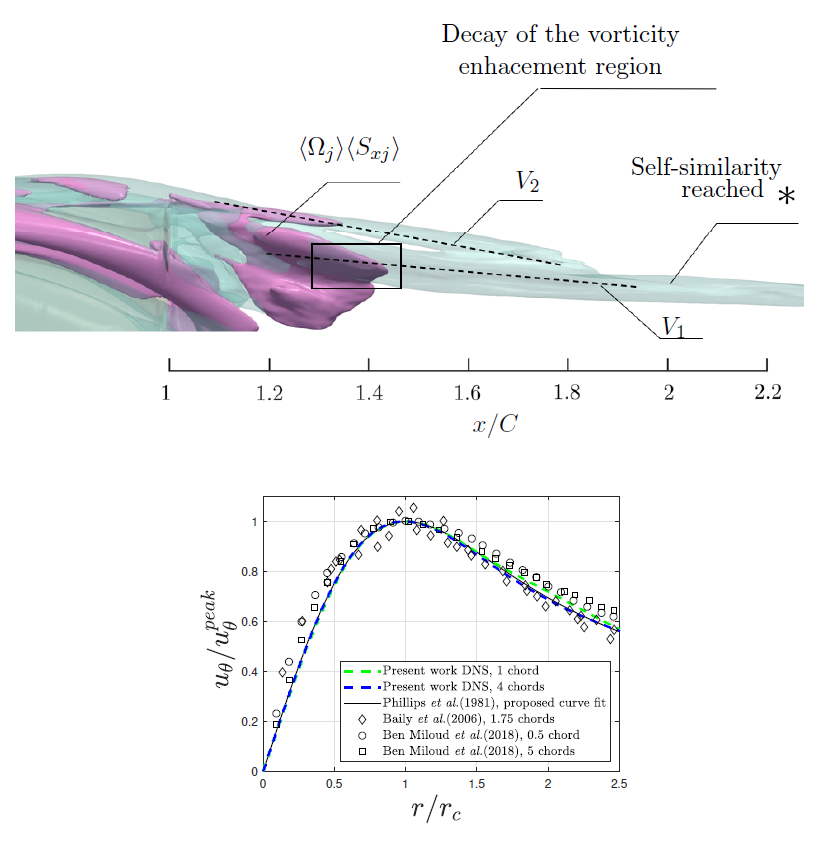}
\caption{\label{fig:stretching_convection_momentum_transfer} Isosurface of the vorticity intensification term  $\langle \Omega_{j}\rangle \langle S_{xj} \rangle =12$ (purple colour) superimposed on $\langle Q \rangle =5$ (top). Normalised azimuthal velocity profile inside V1 at two positions $x/C=2$ and $4$, and comparison with self-similar profiles from the literature (bottom).}
\end{figure*}
Fig.~\ref{fig:stretching_convection_momentum_transfer} shows an isosurface of the total intensification term in the region $x/C=1-2.2$ superposed with the $\langle Q \rangle =5$ to visualise the V1 and V2 vortex structures. V2 has completely merged with V1 at around $x/C=1.8$. Note that the isosurface value of $\langle \Omega_{j}\rangle \langle S_{xj} \rangle=12$ is the same as one plotted in panel (c) of Fig.~\ref{fig:Capture_origins_of_streamwise_vortices_iteration_1}. Once the vorticity intensification term has decayed in the wake (at a location indicated by an asterisk), the azimuthal velocity profile in the radial direction inside V1 becomes self-similar. This is clearly evidenced in the bottom panel, which demonstrates very close agreement with self-similar velocity profiles reported in the literature (\cite{phillips1981turbulent, bailey2006effects}). The matching is very good, both in the inner part where solid-body rotation is found (linear variation), and also in the outer region $r/{r_c}\approx[0.7-2.5]$, where the effect of the potential external flow is felt ($r_c$ indicates the radial location of the maximum azimuthal velocity). Further downstream, the only mechanism affecting V1 is viscous diffusion, thus it decays slowly and persists until the end of the computational domain.   

\section{Turbulence structure of the separated wake} \label{sec:turbulence_characteristics}

\begin{figure*}[h!]
\centering
\includegraphics[width=0.8\linewidth]{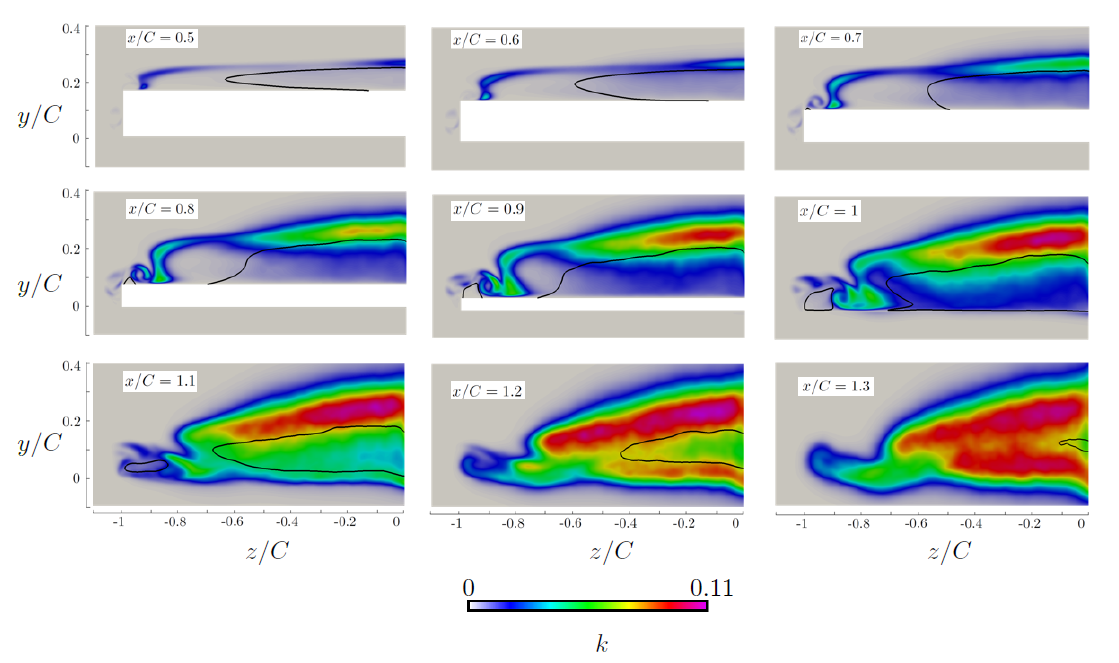}
\caption{\label{fig:k_full_span_umean_contour} Contours of $k$ plotted at equally spaced planes $x/C = 0.5-1.3$. The black contour line represents $\langle {U_1} \rangle =0$.}
\end{figure*}

In this section, we examine the spatial distribution of turbulent kinetic energy (TKE), $k$, and its production. The objective is to identify the most intense regions and uncover the main production mechanisms. We focus on the near-field, i.e.\ distance $5\%-30\%C$ from the trailing edge. In this region, the flow is far from being self-similar, thus well known viscous vortex models, such as the Burgers or Sullivan vortex (\cite{green2012fluid,wu2015vortical}), do not apply. 

The distribution of $k$ across the span at different planes is shown in Fig.~\ref{fig:k_full_span_umean_contour}. Between $x/C=0.5-0.8$ $k$ grows rapidly in the area just above the $\langle {U_1} \rangle =0$ isoline (marked with a solid black line). The high shear in this localised region triggers the formation and growth of Kelvin-Helmholtz instabilities, shown earlier in Fig.~\ref{fig:Q-Criterion_all}. Inside the recirculation bubble the value of $k$ is low. The transition point is located approximately between $x/C \approx 0.8-0.9$. Note that the $k$ contours in the suction side closely follow the shape of the distorted shear layer that carries the RV vortex. Close to the tip, low values of $k$ are encountered. In the near wake at $x/C=1.2$, two regions of high turbulent intensity can be detected, one originates from the suction side and the other from the pressure side, with the former being more widespread.  Further downstream, at $x/C=1.3$, the two regions merge.

\begin{figure*}[h!]
\centering
\includegraphics[width=0.8\linewidth]{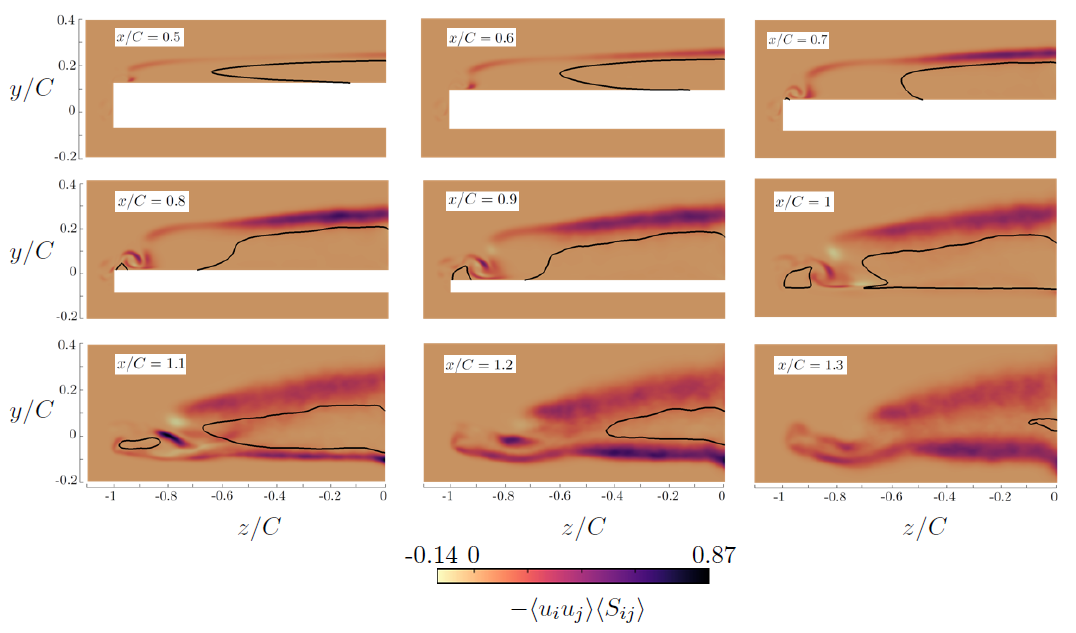}
\caption{Contour plots of TKE production $P=-\langle u_iu_j\rangle \langle S_{ij} \rangle$ at equally spaced planes $x/C=0.5-1.3$. The black contour line represents ${\langle U_1} \rangle=0$.}
\label{fig:k_production_full_span_umean_contour}
\end{figure*}

Fig.~\ref{fig:k_production_full_span_umean_contour} shows contour plots of the production term $P=-\langle u_iu_j\rangle \langle S_{ij} \rangle$ in the same planes. The plots confirm the presence of a high production region just above the recirculation bubble; this explains the shape of $k$ contours shown in the previous Fig.~\ref{fig:k_full_span_umean_contour}. The extent of the high $P$ region in the spanwise direction closely follows the extent of the recirculation bubble, as can be seen in planes $x/C=0.7-1.1$. Further downstream, at $x/C=1.1-1.2$, the high $P$ region envelops the shrinking bubble. The mixing in the wake fills the regions of low $P$ with substantial values of $k$, as can be more clearly seen at plane $x/C=1.3$ (compare with previous figure \ref{fig:k_full_span_umean_contour}). 

The above broad brush picture needs to be complemented with some additional details. For example, notice the localised pockets of high production in the area between the two recirculation zones. These pockets first appear at $x/C=0.8$ but can be detected even at $x/C=1.3$. They can explain the local overshoot of $k$, visible in Fig.~\ref{fig:k_full_span_umean_contour} at all planes between $x/C=0.8-1.2$, at the spanwise location $z/C \approx -0.8$. Further outboard, at $z/C=-0.9$, where the centre of the fully formed  wing tip  vortex V1 is located (refer to Fig.~ \ref{fig:Vortex_stretching_component_evolution_iteration_1}) both $k$ and $P$ attain very low values.

\begin{figure*}[h!]
\centering
\includegraphics[width=0.7\linewidth]{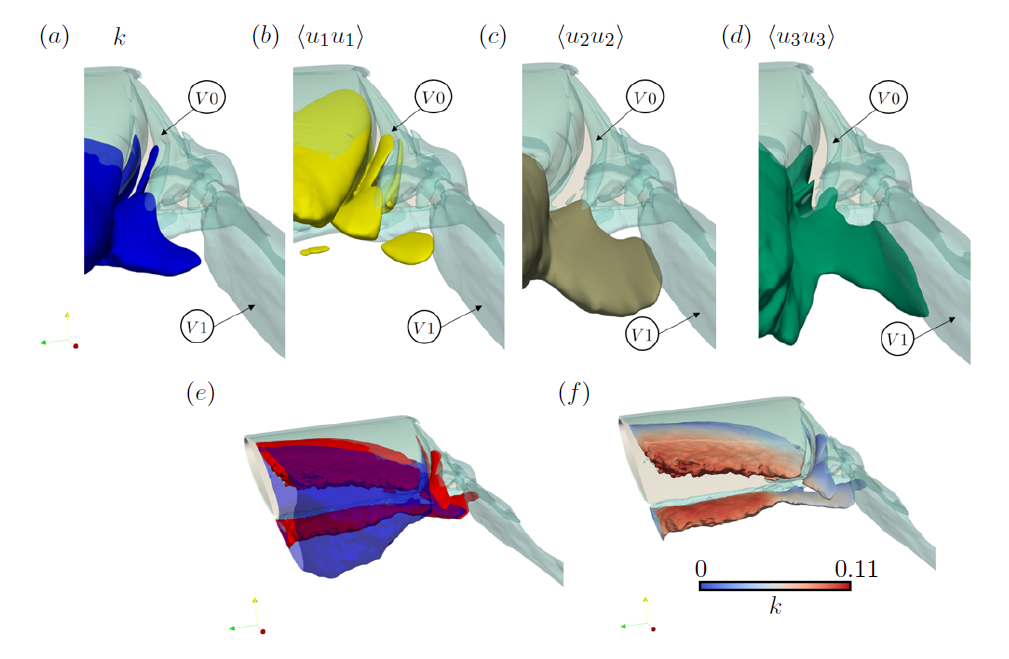}
\caption {Isosurfaces of $(a)$ $k$, $(b)$ $\langle u_{1}u_{1} \rangle$, $(c)$ $\langle u_{2}u_{2} \rangle$ and $(d)$ $\langle u_{3}u_{3}\rangle$ with values $0.05$, $0.05$, $0.04$ and $0.02$ respectively. These values correspond to $(a)$ $\approx45\%$, $(b)$ $\approx55\%$, $(c)$ $\approx49\%$, and $(d)$ $\approx29\%$ of the maximum value of the corresponding quantities at the z-y plane located at x/C=1.2. (e) Isosurfaces of $k$ with value $0.5$ (blue) and turbulence production with value $0.2$ (red).  The isosurface values correspond to $\approx55\%$ and $\approx31\%$ of the maximum value of $k$ and $\mathcal{P}$ respectively in the z-y plane located at x/C=1.2. (f) Turbulence production $\mathcal{P}$ colored by $k$. The isosurface $\langle Q \rangle =5$ is superimposed in all plots to visualise the vortical structures.}
	\label{fig:Re_yy_power_point_arragements_iterations_1}
\end{figure*}

The increased turbulent activity between the two recirculation bubbles can be also detected in the Reynolds stresses isosurface plots shown in Fig.~\ref{fig:Re_yy_power_point_arragements_iterations_1}. Analysis shows that $\langle u_{1}u_{1} \rangle$ and $\langle u_{2}u_{2} \rangle$ dominate over $\langle u_{3}u_{3}\rangle$. It can be seen that $k$ and $\langle u_{1}u_{1} \rangle$ isosurfaces wrap around the outer boundary of V0 and the large bubble originating from the leading edge. The levels of turbulent production and $k$ are small in the V0 core, most likely due to the stabilizing effect of solid body rotation   (\cite{zeman1995persistence}). 

\subsection{Production mechanisms of Reynolds stresses}

We now proceed to study the production terms of the Reynolds stresses of $\langle u_{1}u_{1}\rangle$ and  $\langle u_{2}u_{2}\rangle$, given by \cite{pope2000turbulent},
\begin{eqnarray}
	\mathcal{P}^{R-s}_{11}=-2\left[ \langle u_{1}u_{1}\rangle  \frac{\partial \langle U_1 \rangle}{\partial x_{1}}   +\langle u_{1}u_{2}\rangle  \frac{\partial \langle U_1 \rangle}{\partial x_{2}} +\langle u_{1}u_{3}\rangle  \frac{\partial \langle U_1 \rangle}{\partial x_{3}} \right] 
	\label{eq:P_11}
\end{eqnarray}
\begin{eqnarray}
	\mathcal{P}^{R-s}_{22}=-2\left[\langle u_{2}u_{1}\rangle  \frac{\partial \langle U_2 \rangle}{\partial x_{1}}   +\langle u_{2}u_{2}\rangle  \frac{\partial \langle U_2 \rangle}{\partial x_{2}} +\langle u_{2}u_{3}\rangle  \frac{\partial \langle U_2 \rangle}{\partial x_{3}} \right] 
	\label{eq:P_22}
\end{eqnarray}

The magnitude of $\langle u_{3}u_{3}\rangle$ (and its production) is smaller in comparison with the other two stresses and is not investigated.

\begin{figure*}[h!]
\centering
\includegraphics[width=0.8\linewidth]{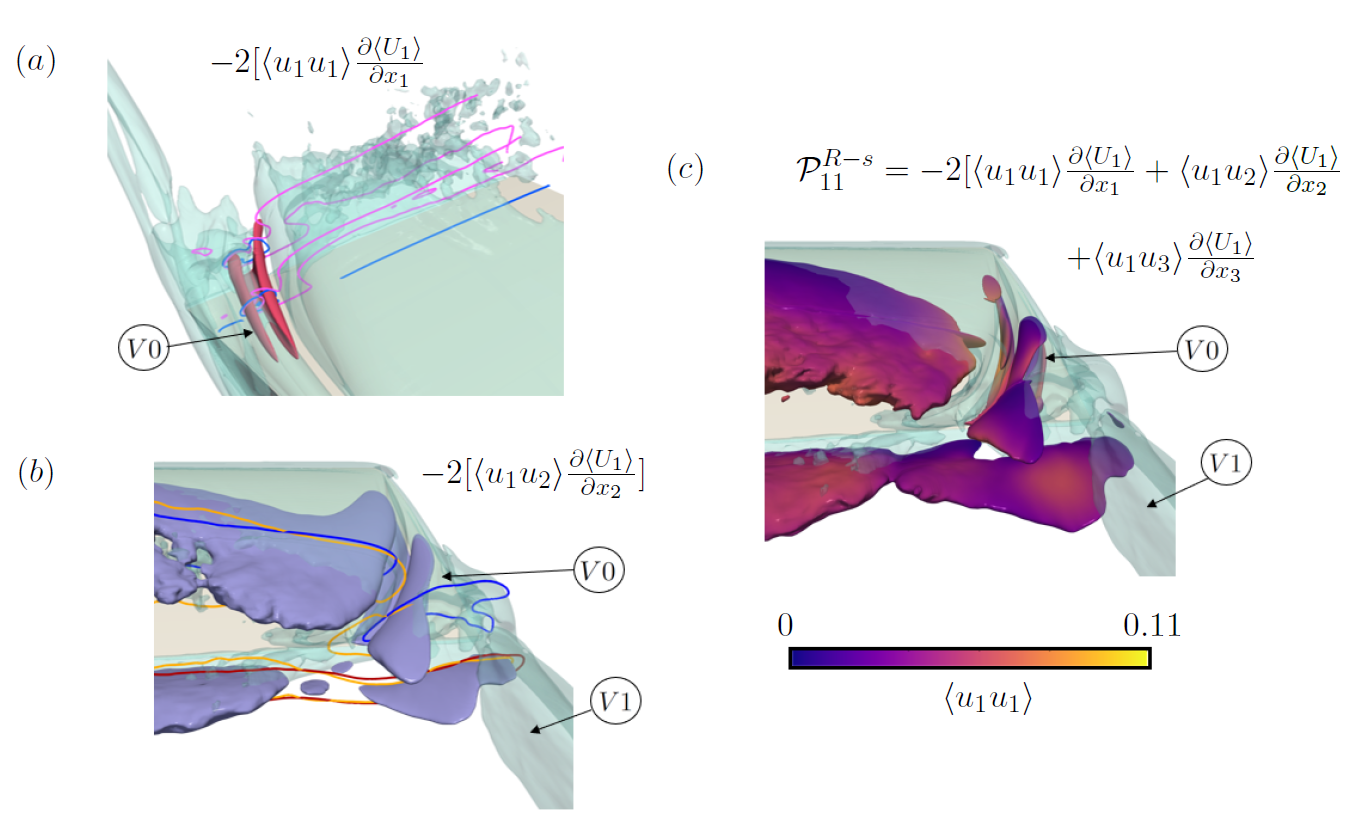}
\caption{\label{fig:u1u1_production_iteration_1} Production terms of $\langle u_{1}u_{1}\rangle$ (a) The magenta isosurface represents $-2[\langle u_{1}u_{1}\rangle  \frac{\partial \langle U_1 \rangle}{\partial x_{1}}]=0.4$ with pink lines representing the contours of $\langle u_{1}u_{1}\rangle=0.03$ and blue line representing compression $\frac{\partial \langle U_1 \rangle}{\partial x_{1}}=-2$. (b) The purple isosurface represents $-2[\langle u_{1}u_{2}\rangle  \frac{\partial \langle U_1 \rangle}{\partial x_{2}}]=0.5$, blue lines represent the contours of shear strain coming from the top wing surface with $\frac{\partial \langle U_1 \rangle}{\partial x_{2}}=-7$ and orange lines $\langle u_{1}u_{2}\rangle=0.01$, the red line represents shear from the bottom wing surface with $\frac{\partial \langle U_1 \rangle}{\partial x_{2}}=7$ and the corresponding orange line $\langle u_{1}u_{2}\rangle=-0.01$ and (c) total production of $\langle u_{1}u_{1}\rangle$: iso-surface $\mathcal{P}^{R-s}_{11}=0.3$ colored by $\langle u_{1}u_{1}\rangle$. The isosurface values are equal to $(a)$ $\approx45\%$, $(b)$ $\approx36\%$ and $(c)$ $\approx27\%$ of the positive maximum value of the corresponding quantities in the z-y plane located at $x/C=1.2$ (at $x/C=0.9$ for $(a)$). The isosurface $\langle Q \rangle =5$ is superimposed in all plots to visualise the vortical structures.}
\end{figure*}

\begin{figure*}[h!]
\centering
\includegraphics[width=0.8\linewidth]{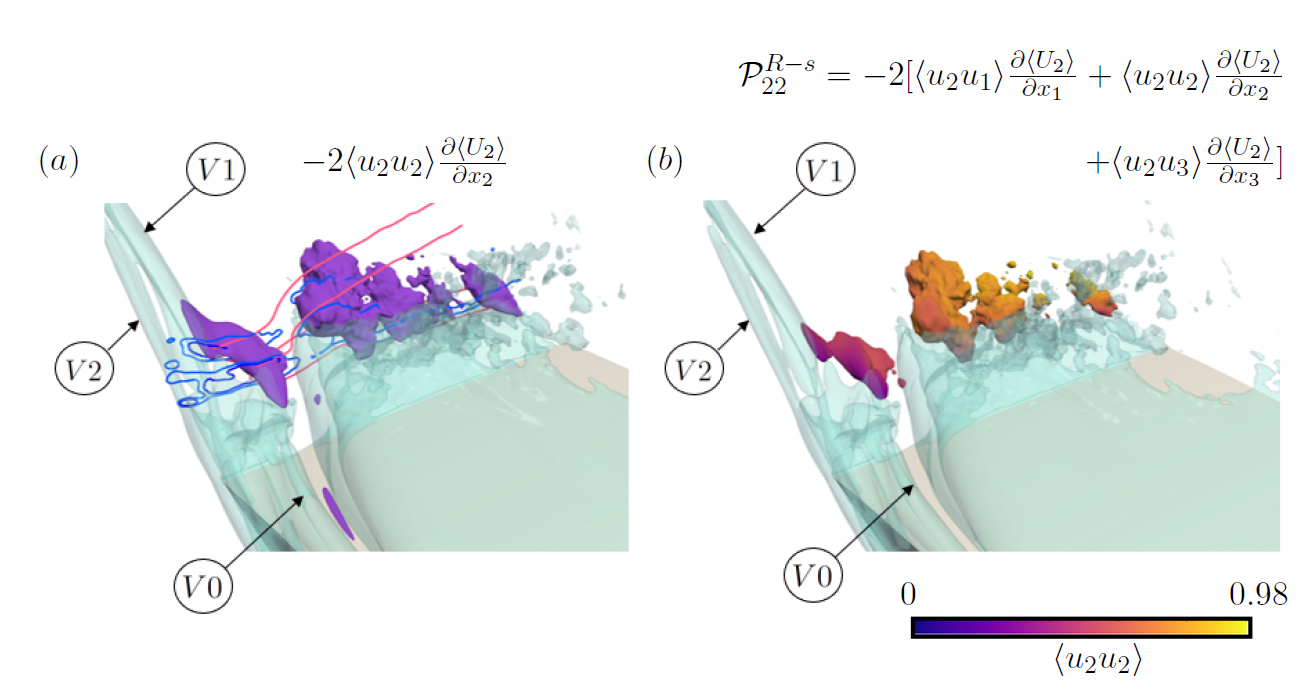}
\caption{\label{fig:u2u2_production_iteration_1} Production terms of $\langle u_{2}u_{2}\rangle$ (a) Isosurface of $-2\langle u_{2}u_{2}\rangle  \frac{\partial \langle U_2 \rangle}{\partial x_{2}}=0.3$, blue lines represent contours  $\frac{\partial \langle U_2 \rangle}{\partial x_{2}}=-3,-4$ and the pink line represents $\langle u_{2}u_{2}\rangle=0.04$ (b) Isosurface of 	$\mathcal{P}^{R-s}_{22}=0.3$ colored by $\langle u_{1}u_{1}\rangle$. The contour values correspond to $(a)$ $\approx61\%$ and $(b)$ $\approx64\%$ of their respective positive maximum value in the z-y plane located at x/C=1.2. The isosurface $\langle Q \rangle =5$ is superimposed in all plots to visualise the vortical structures.}
\end{figure*}

Figures~\ref{fig:u1u1_production_iteration_1} and  \ref{fig:u2u2_production_iteration_1} show isosurfaces of  $\mathcal{P}^{R-s}_{11}$ and $\mathcal{P}^{R-s}_{22}$ respectively. In each plot, isosurfaces of the most important components that make up the production terms according to \eqref{eq:P_11} and \eqref{eq:P_22} are also plotted. Fig.~\ref{fig:u1u1_production_iteration_1} indicates that the term $-2\langle u_{1}u_{2}\rangle  \frac{\partial \langle U_1 \rangle}{\partial x_{2}}$ is by far the largest contributor of $\mathcal{P}_{11}$, compare panels (b) and (c). This is of course expected since both the separating shear layer on the suction side and the attached flow in the pressure side are associated with strong shear $\frac{\partial \langle U_1 \rangle}{\partial x_{2}}$ (note however that the production is stronger in the suction side because flow separation results in higher shear). It is also interesting to notice the localised contribution of the normal term, $-2\langle u_{1}u_{1}\rangle  \frac{\partial \langle U_1 \rangle}{\partial x_{1}}$, which wraps around vortex V0, as can be seen in panel (a).

The production term $\mathcal{P}^{R-s}_{22}$ is shown in panel (b) of Fig.~\ref{fig:u2u2_production_iteration_1}. The dominant component is $-2\langle u_{2}u_{2}\rangle  \frac{\partial \langle U_2 \rangle}{\partial x_{2}}$, which is shown in panel (a). The two other components are not included because their contribution is negligible; this can be attested by the resemblance of the isosurfaces in the two panels. The production term is maximised downstream of the trailing edge, where higher local values of negative normal strain $\frac{\partial \langle U_2 \rangle}{\partial x_{2}}$ are encountered.  Assuming that $\frac{\partial \langle U_3 \rangle}{\partial x_{3}}$ is small, from the continuity equation we get $\frac{\partial \langle U_2 \rangle}{\partial x_{2}}=-\frac{\partial \langle U_1 \rangle}{\partial x_{1}}$, so the regions of strong $-2\langle u_{2}u_{2}\rangle \frac{\partial \langle U_2 \rangle}{\partial x_{2}}$ are collocated with the areas of rapid streamwise velocity recovery, i.e.\ large  $\frac{\partial \langle U_1 \rangle}{\partial x_{1}}$ acceleration (or normal strain rate). This is the area shortly downstream of the closure of the recirculation region close to the tip.

\begin{figure*}[h!]
\centering
\includegraphics[width=0.8\linewidth]{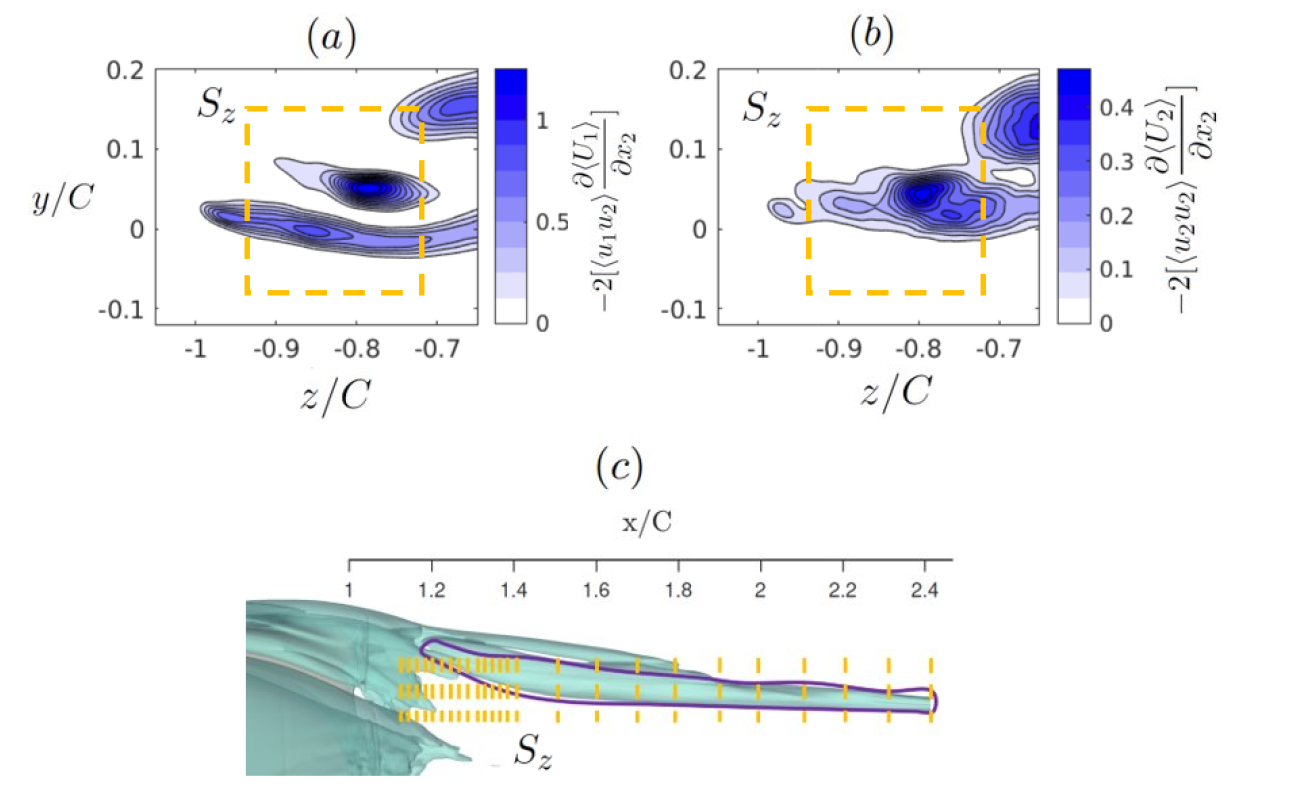}
	\caption{(a) Contours of $-2[\langle u_{1}u_{2}\rangle  \frac{\partial \langle U_1 \rangle}{\partial x_{2}}]$  (b) $-2[\langle u_{2}u_{2}\rangle  \frac{\partial \langle U_2 \rangle}{\partial x_{2}}]$ superimposed with the integration area ${S_z}$ at $x/C=1.2$. (c) Streamwise locations of the integration planes superimposed on $\langle Q \rangle =5$ isosurface.}
	\label{fig:control_volume_integration}
\end{figure*}

In order to probe further the interaction between the elongation rate,  $\langle S_{xx} \rangle$, vortex stretching, circulation and production of turbulent kinetic energy in the near wake, we integrate the aforementioned quantities in a number of  cross-steam planes from $x/C=1.1-2.4$.  The boundary of the  integration area ${S_z}$ is shown with a closed dashed line in panels (a) and (b) of Fig.~\ref{fig:control_volume_integration},  where contours of $-2\langle u_{1}u_{2}\rangle  \frac{\partial \langle U_1 \rangle}{\partial x_{2}}$ and $-2\langle u_{2}u_{2}\rangle  \frac{\partial \langle U_2 \rangle}{\partial x_{2}}$ are plotted respectively at plane $x/C=1.2$. The locations of the planes themselves are marked in Fig.~\ref{fig:control_volume_integration}(c). The three areas of large production $-2\langle u_{1}u_{2}\rangle  \frac{\partial \langle U_1 \rangle}{\partial x_{2}}$ in panel (a)  correspond to flow originating from the pressure side (bottom horizontal region), the suction side (top right area) and between the two recirculation zones (middle region), see also contours of the total production $P=-\langle u_iu_j\rangle \langle S_{ij} \rangle$ at plane $x/C=1.2$ at Fig.~\ref{fig:k_production_full_span_umean_contour}. The patches of large $-2\langle u_{2}u_{2}\rangle  \frac{\partial \langle U_2 \rangle}{\partial x_{2}}$ in panel (b) correspond to the two areas shown in Fig.~\ref{fig:u2u2_production_iteration_1}(a).

\begin{figure*}[h!]
\centering
\includegraphics[width=0.76\linewidth]{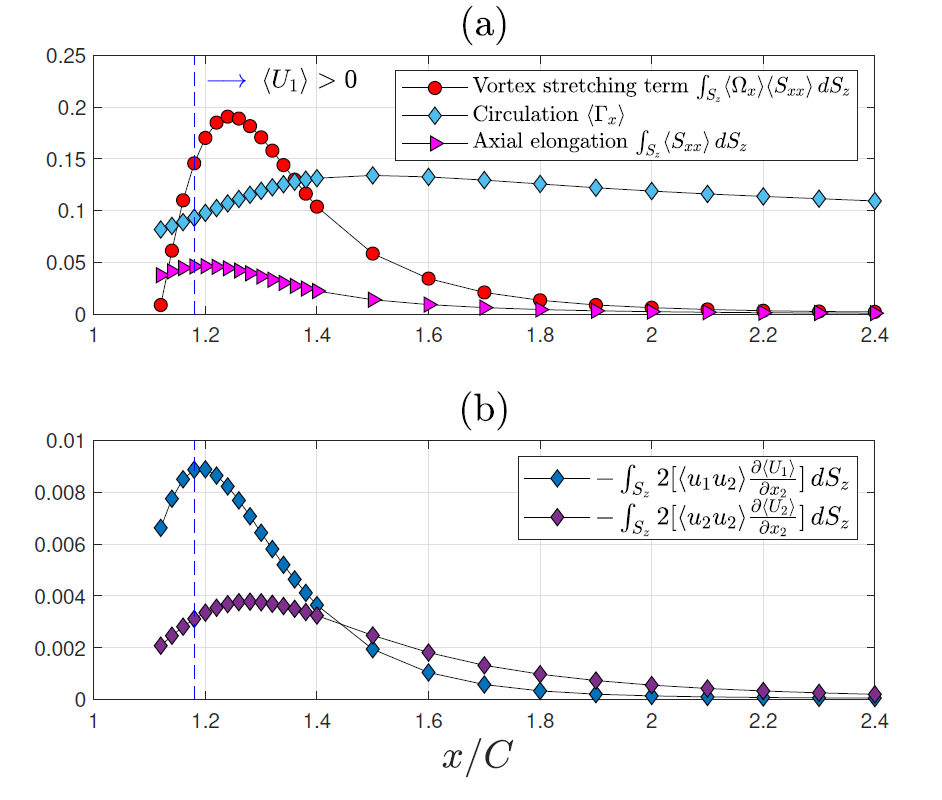}
  \caption{Streamwise evolution of quantities related to (a) the vorticity field and (b) the production of Reynolds stresses. The quantities shown are integrated over the surface ${S_z}$ marked in panels (a) and (b) of Fig.~\ref{fig:control_volume_integration}. The vertical dashed line indicates the streamwise location where $\langle U_1 \rangle=0$.}
\label{fig:Integration}
\end{figure*}

In panel (a) of Fig.~\ref{fig:Integration} we plot the variation along $x/C$ of the surface integrals of vortex stretching, axial elongation and production terms over ${S_z}$. We also superimpose the circulation $\langle \Gamma_x \rangle$ (surface integral of $\langle \Omega_x \rangle$ ). 
This panel shows that the peaks of $\int_{S_z} \langle \Omega_x \rangle \langle S_{xx} \rangle\,dS_z $, $\langle \Gamma_x \rangle$  and $\int_{S_z}\langle S_{xx} \rangle\,dS_z$ are not in phase. The axial elongation term peaks at the streamwise location where $\langle U_1 \rangle=0$ (marked with the vertical dashed line). This is the result of flow recovery, as the flow comes out of the recirculation region around the tip. The vortex stretching term peaks slightly further downstream. The latter term causes the growth of circulation,  $\langle \Gamma_x \rangle$ which peaks at about $x/C \approx 1.5$.  Note the very weak decay of $\langle \Gamma_x \rangle$ with respect to the other two terms. The only mechanism responsible for the decay for  $x/C > 2.0$ is viscous diffusion, which is a slow term and explains the persistence of V1 until the exit of the computational domain. Panel (b) shows that the production component due to shear $-\langle u_{1}u_{2}\rangle  \frac{\partial \langle U_1 \rangle}{\partial x_{2}}$ peaks earlier compared to the normal component $-2\langle u_{2}u_{2}\rangle  \frac{\partial \langle U_2 \rangle}{\partial x_{2}}$ and it reaches a higher value. The peak of the former term almost coincides with the location where $\langle U_1 \rangle=0$. Note also the faster decay rate of $-2\langle u_{1}u_{2}\rangle  \frac{\partial \langle U_1 \rangle}{\partial x_{2}}$ compared to $-2\langle u_{2}u_{2}\rangle  \frac{\partial \langle U_2 \rangle}{\partial x_{2}}$. While circulation is (almost) constant, both production terms decay because of the solid body rotation velocity profile within V1, see bottom row of figure \ref{fig:stretching_convection_momentum_transfer}. The fast decay of $-2\langle u_{1}u_{2}\rangle  \frac{\partial \langle U_1 \rangle}{\partial x_{2}}$ between $x/C \approx 1.2-1.4$ is probably due to the flow recovery away from the recirculation region that smooths the streamwise velocity in the y direction.

\section{Conclusions}\label{sec:conclusions}
DNS of a half-span NACA 0018 wing with square wingtip profile was performed at a $Re_c=10^4$ and $10^\circ$ angle of attack. A thorough analysis of the main vorticity and turbulent production mechanisms above the wing and in the near wake was carried out. The flow was found to separate close to the leading edge and did not reattach on the suction side of the wing. Close to the leading edge, the separating region occupied most of the span, but once the secondary flow in the wingtip started to develop, the left boundary of the recirculating bubble was displaced inboard. In the proximity of the tip, the flow remained attached, but another recirculation zone appeared close to the trailing edge. 

In total 5 vortical structures were identified, with one of them rotating in the counter-clockwise direction. The underlying mechanism was traced to the low pressure region in the flat wingtip close to the leading edge, which created a spanwise motion in the outboard direction. This low pressure was associated with the formation of a small recirculation zone, arising from the flow separation at the sharp wingtip  around the leading edge. 


We analysed the terms of the vorticity transport equation and showed how each of the identified vortices sustained or lost their strength as they approached the trailing edge, with particular focus on the role of vortex stretching/compression and tilting. Different mechanisms were found to be at play in the separating shear layers and in the vortex cores. The emergence of the main wingtip vortex V1 from the amalgamation and interaction of the different vortices was also elucidated. The presence of the recirculation zone close to the trailing edge resulted in the distortion of the shear layers originating from the pressure side.

The distribution of turbulent kinetic energy around the main wing tip vortex $V1$ was found to be highly asymmetric and high values were concentrated on the right of its axis. Apart from the main separating shear layer, patches of strong production were also found in the area between the two recirculation bubbles in the near wake.

Analysis of the production terms of individual Reynolds stresses has enabled us to uncover the main production mechanisms. It was found that $-2\langle u_{1}u_{2}\rangle  \frac{\partial \langle U_1 \rangle}{\partial x_{2}}$  is one of the dominant terms responsible for maintaining high levels of $\langle u_{1}u_{1}\rangle$. This is the standard production term for wall-bounded flows. Additionally, the wall-normal stress source term $-2\langle u_{2}u_{2}\rangle  \frac{\partial \langle U_2 \rangle}{\partial x_{2}}$ was responsible for maintaining $\langle u_{2}u_{2}\rangle$ primarily in the spiral wakes in the near field. 

The vortex stretching rate, axial elongation, streamwise vorticity and production terms were integrated over square y-z surfaces surrounding the V1 vortex, and plotted against the streamwise direction $x/C$. When the stretching rate became negligible, the circulation (integral of streamwise vorticity) reached a peak and then started to decay very slowly, explaining the persistence of V1 until the end of the computational domain. The production terms become negligible at $x/C \approx 2$ as expected, because the velocity profile within the core of V1 (solid body rotation) cannot sustain turbulence.  

The detailed analysis presented aims to  contribute to a better understanding of separated vortical flows and can hopefully lead to actuation strategies to manipulate them.

\subsection*{Acknowledgements}
The authors are grateful to the ARCHER2 UK National Supercomputing Service (https://www.archer2.ac.uk), with access via the UK Turbulence Consortium (EPSRC grant EP/R029326/1). Additional computational resources were provided by the UK Materials and Molecular Modelling Hub, which is partially funded by EPSRC (EP/P020194/1 and EP/T022213/1) and Imperial College London. The first author would like to acknowledge the financial support by the National Program of Scholarships and Educational Credit (PRONABEC) of Peru as part of the Beca Generación Bicentenario PhD scholarship.

\bibliographystyle{unsrtnat}
\bibliography{references}  






\end{document}